\documentclass[times,preprint,3p,sort&compress]{elsarticle}  
\usepackage{amssymb}
\usepackage{amsmath}
\usepackage{lipsum}
\usepackage{mathtools,cuted}

\usepackage{graphicx}  
\usepackage{bm}        
\usepackage{amssymb}   
\usepackage{amsbsy}
\usepackage{amsmath}
\usepackage{amsfonts}
\usepackage{color}
\usepackage[dvipsnames]{xcolor}
\usepackage{amsthm}
\usepackage{graphics}
\usepackage{graphicx}
\usepackage{xcolor}
\usepackage{caption}
\usepackage{subcaption}
\usepackage{amssymb}
\usepackage{enumitem}
\setlist[itemize]{leftmargin=*}

\usepackage{xcolor}
\usepackage{graphicx}

\usepackage[hidelinks]{hyperref}
\hypersetup{
  colorlinks   = true, 	
  urlcolor     = blue, 	
  linkcolor    = blue, 	
  citecolor   = red 	
}

\DeclareMathOperator{\sech}{sech}
\DeclareMathOperator{\arctanh}{arctanh}
\DeclareMathOperator{\arccosh}{arccosh}

\begin{document}
\begin{frontmatter}

\title{Wobbling kinks and shape mode interactions  in a coupled two-component $\phi^4$ theory}
\author[salamanca1,salamanca2]{A. Alonso-Izquierdo}\ead{alonsoiz@usal.es}
\author[valladolid]{D.~Migu\'elez-Caballero}
\ead{david.miguelez@uva.es}
\author[valladolid]{L.M. Nieto\corref{cor1}}
\ead{luismiguel.nieto.calzada@uva.es}

\cortext[cor1]{Corresponding author}

\address[salamanca1]{Departamento de Mat\'ematica Aplicada, Universidad de Salamanca, Casas del Parque 2, 37008, Salamanca, Spain}
\address[salamanca2]{IUFFyM, Universidad de Salamanca, Plaza de la Merced 1, 37008, Salamanca, Spain}
\address[valladolid]{Departamento de F\'{\i}sica Te\'{o}rica, At\'{o}mica y \'{O}ptica,
Universidad de Valladolid, 47011 Valladolid, Spain}

\begin{abstract}
The dynamics of a wobbling kink   in a two-component  coupled $\phi^4$ scalar field theory (with an excited orthogonal shape mode)  is addressed. 
For this  purpose, the vibration spectrum of  the second order
small kink fluctuation is studied in order to find the corresponding vibration modes associated  to the first (longitudinal) and second (orthogonal) field components. 
By means of this analysis, it was found that the number of possible shape modes depends on the value of the coupling constant. 
It is notable that when one  of the orthogonal field shape modes  is initially triggered, the unique shape mode of the longitudinal field is also activated. 
This coupling causes the kink to emit radiation with twice the frequency of excited mode in the first field component. Meanwhile,  in the orthogonal channel we find radiation with two different frequencies: one is three times the frequency of the orthogonal wobbling mode and another is the sum of the frequencies of the longitudinal shape mode and the triggered mode. 
All the analytical results obtained in this study have been successfully contrasted with those obtained through numerical simulations.
\end{abstract}


\end{frontmatter}

\section{Introduction}

It is well known that topological defects have played a key role in the last decades to understand nonlinear phenomena in many areas, such as condensed matter \cite{Bishop1978, Rajaraman1982,Dauxois2006,CuevasMaraver2014},  cosmology \cite{Kibble1976, Vachaspati2006,Vilenkin1994},  superconductivity \cite{Buzea1998} or quantum communications \cite{Dymarsky2021,Buican2023}. 
Among all the existing range of topological defects, the kinks, which arise when considering nonlinear scalar field theories, are undoubtedly the simplest.
In this context, the most famous theory is the $\phi^4$ model, for which kink scattering and collisions have been extensively investigated due to the complex pattern of collisions in which a fractal pattern arises \cite{Goodman2005, AlonsoIzquierdo2021b,AlonsoIzquierdo2021c,Mohammadi2022,Kevrekidis2019}. 
This same type of phenomena has also been observed when studying other more complex scalar models, such as theories with other polynomial potentials such as $\phi^6$ \cite{Springer2019,Dorey2011,Marjaneh2017,Gani2014} or $\phi^8$ \cite{Belendryasova2018, Bazeia2023, Gani2015}, theories with two-component scalar fields \cite{AlonsoIzquierdo2021,AlonsoIzquierdo2023, Katsura2013, Bazeia1995, Shifman1998, Ashcroft2016, AlonsoIzquierdo2019, Aguirre2020, AlonsoIzquierdo2002, Halavanau2012}, or more complex ones \cite{Blinov2022, Christov2021, Askari2020, Takyi2023, Mohammadi2021}. 
In addition to numerical analysis methods,  analytical tools have been developed to study and understand the  physical behavior of these fascinating solutions to the $\phi^6$ and $\phi^8$ models. An example of what we just said can be found in the use of the  moduli space approximation \cite{Sugiyama1978, NavarroObregon2023, Manton2004, Adam2022, Adam2023, Takyi2016} to reduce the degrees of freedom of the system.  These methods have successfully  replicated phenomena observed in kink solutions, such as radiation emission or oscillon formation.

When kink scattering phenomena are studied, among the possible final configurations of the system it is possible to find \textit{wobbling kinks} or \textit{wobblers}, which  consists of a kink solution where  one of its vibration modes has been triggered due to the energy transfer mechanism between its shape modes and the kinetic energy of the kink. 
In this context, some perturbative approaches have been developed to  try to better understand how a wobbler behaves. 
It was found  \cite{Barashenkov2019, Barashenkov2009,Manton1997} that a wobbling kink in the $\phi^4$ model emits radiation with twice the frequency associated with its shape mode, which also causes a decay in the wobbling amplitude due to the loss of energy in the form of radiation.   
These perturbative theories have also been  implemented to study the evolution of wobbling kinks in two-component scalar field theories, such us the MSTB model \cite{AlonsoIzquierdo2023}. 
In the present article, the theory under study will consist of two separate copies of a $\phi^4$ theory coupled by a cross term \cite{Halavanau2012}. 
The main difference between this theory and the MSTB model is that in the first case the shape mode structure is more complex than in  the second, since the number of shape modes corresponding to the second field depends on the coupling constant between the two copies of the $\phi^4$ model.

This paper is structured as follows: in Section~\ref{Section2}  the model under study will be presented and kink solutions will be found. In addition, the linear stability of these solutions will be analyzed, which will allow us to find the corresponding vibration eigenfunctions and eigenfrequencies. 
In Section~\ref{Section3} the perturbative approach introduced by Manton and Merabet in~\cite{Manton1997} will be used to describe the behavior of a kink when one of its shape modes is activated. In Section~\ref{Section4} all the analytical results found in the preceding sections will be compared with those obtained by numerical simulations. The paper ends with a some concluding remarks.

\section{The two-component coupled $\phi^4$ model: kink solutions, linear stability and shape modes }\label{Section2}

As already mentioned above, in the following sections we will deal with a two-component real scalar field theory, consisting of two separate copies of a $\phi^4$ model, coupled by means of a cross term $\kappa \phi^2 \psi^2$, where $\kappa$ is  a real positive parameter.  Thus, the dynamics of this physical system is governed by the Lagrangian density
\begin{equation}\label{LagrangianDensity}
    \mathcal{L}=\frac{1}{2}\partial_\mu \phi \partial^\mu \phi+\frac{1}{2}\partial_\mu \psi \partial^\mu \psi- U(\phi, \psi),
\end{equation}
where the potential $U(\phi,\psi)$ is given by 
\begin{equation}\label{Potential}
    U(\phi, \psi)=\frac{1}{2}(\phi^2-1)^2+\frac{1}{2}(\psi^2-1)^2+ \kappa \phi^2 \psi^2 -\frac{1}{2} ,
\end{equation}
which is completely symmetric  in the interchange of $\phi$ for $\psi$.
In equations \eqref{LagrangianDensity}--\eqref{Potential} we assume that $\phi$ and $\psi$ are real scalar fields and the Minkowski metric is taken in the usual form:  $g_{\mu,\nu}=\text{diag} \{ 1,-1\}$. 
In general, the field equations that govern the evolution of both fields,  which can be  calculated from the Lagrangian density \eqref{LagrangianDensity}, are
\begin{eqnarray}
  &&  \partial_{tt} \phi -\partial_{xx} \phi + 2 \phi (\phi^2 -1 + \kappa \psi^2)=0, \label{FieldEqn1} \\ 
   && \partial_{tt} \psi -\partial_{xx} \psi + 2 \psi (\psi^2 -1 + \kappa\phi^2)=0. \label{FieldEqn2}
\end{eqnarray}

\subsection{Vacua and kinks}

Both the vacuum structure (the set of minima of the potential that make it take the value zero) and the kinks of the model depends on the value of the parameter $\kappa$, as explained below.

\begin{itemize}

\item[$\rhd$]
When $\kappa<1$, the  vacua are the following four constant solutions of the field equations \eqref{FieldEqn1}--\eqref{FieldEqn2}
\begin{equation}
\mathcal{M}_{\kappa<1}=\left\{ 
\frac1{\sqrt{1+\kappa}} 
 \left( 
\begin{array}{c}
(-1)^a  \\ 
(-1)^b
\end{array}
\right) 
, \  a,b=0,1\right\}.
\label{vacua4<1}
\end{equation}
In fact, for this last set of solutions to be strictly vacua, it is necessary to add to the potential \eqref{Potential} a constant  term $\frac{1-\kappa}{2(1+\kappa)}$, so that the potential must really be $U'(\phi,\psi)=U(\phi,\psi)+\frac{1-\kappa}{2(1+\kappa)}$.

The kink structure for  $\kappa<1$ has already been studied in \cite{Halavanau2012}, where the authors investigate the kink scattering characteristics in this model. 
The discrete symmetries of the Lagrangian density \eqref{LagrangianDensity} can be used to find that the equation that governs both components of the kink $K(x)= \begin{pmatrix}
\phi(x)  \\ 
\psi(x)
    \end{pmatrix}
$
is 
\begin{equation}
     -\partial_{xx} \phi^\kappa_K + 2 \phi^\kappa_K \left( (1+\kappa)(\phi_K^\kappa)^2 -1 \right)=0,
\end{equation}
whose solutions, except for an irrelevant translation in the $x$ coordinate, are $ \phi_K^\kappa(x)= \pm  \frac{ \phi_K(x)}{\sqrt{1+\kappa}}$. Here, $\phi_K(x)$ satisfies the well-known field equation of the $\phi^4$ model 
\begin{equation}\label{fi4}
     -\partial_{xx} \phi_K + 2 \phi_K (\phi_K^2 -1 )=0,
\end{equation}
whose solutions are $\pm \phi_K(x)=\pm \tanh (x)$.
Therefore, there are two pairs of different kinks, in such a way that each pair joins two non-adjacent vacua of those we have in \eqref{vacua4<1}, one of them in one direction and the other in the opposite direction, and these four kinks $K(x)$ are of the form
\begin{equation}
    K^{(a,b)} (x)=\frac{\tanh x }{\sqrt{1+\kappa}} \left(
    \begin{array}{c}
    (-1)^a\\ 
    (-1)^b 
    \end{array}
    \right), \quad a,b=0,1.
\end{equation}

\item[$\rhd$]
 When $\kappa>1$, the four vacua solutions are slightly different than in the previous case \eqref{vacua4<1}:
\begin{equation}
\mathcal{M}_{\kappa>1}=
\left\{ 
\left( 
\begin{array}{c}
(-1)^a \\
0 
\end{array}
\right) 
, \ 
\left( 
\begin{array}{c}
0 \\
(-1)^b
\end{array}
\right) 
, \   a,b=0,1\right\}.
\label{vacua4>1}
\end{equation}
It is now possible to find the kink solutions $K(x)= \begin{pmatrix}
\phi(x)  \\ 
\psi(x)
    \end{pmatrix}
$ by cancelling one of its components since, by doing this, from the field equations \eqref{FieldEqn1}--\eqref{FieldEqn2} we can deduce that the evolution of the non-zero component of the field is again described by \eqref{fi4},
whose solutions have been already given.
Taking into account all the previous comments, we can finally infer that for $\kappa>1$ the kinks  in this case are given by the two pairs
\begin{eqnarray}
 K^{(\pm)}_{1} (x) \!\!&\!\! = \!\!&\!\!
 \left(
    \begin{array}{c}
\pm \phi_K(x) \\
0
    \end{array}
    \right)=
     \left(
    \begin{array}{c}
\pm \tanh x \\
0
    \end{array}
    \right)
    ,\label{Kink1}
    \\ 
K^{(\pm)}_{2} (x) \!\!&\!\! =  \!\!&\!\!
 \left(
    \begin{array}{c}
0  \\
\pm\phi_K(x)
    \end{array}
    \right)
=
 \left(
    \begin{array}{c}
0  \\
\pm \tanh x
    \end{array}
    \right).
    \label{Kink2}
\end{eqnarray}
As before, each pair of kinks joins two of the non-contiguous vacua that we have in \eqref{vacua4>1}, one of them in one direction and the other in the opposite. For example, as $x$ goes from $-\infty$ to $+\infty$, $K^{(+)}_{1} (x)$ connects 
$\begin{pmatrix}
-1 \\
0
    \end{pmatrix}
$
with 
$\begin{pmatrix}
1 \\
0
    \end{pmatrix}
$,
while $K^{(-)}_{1} (x)$ connects 
$\begin{pmatrix}
1 \\
0
    \end{pmatrix}
$
with
$\begin{pmatrix}
-1 \\
0
    \end{pmatrix}
$.

\end{itemize}

\begin{figure}[htb] 
         \includegraphics[width=0.48\textwidth]{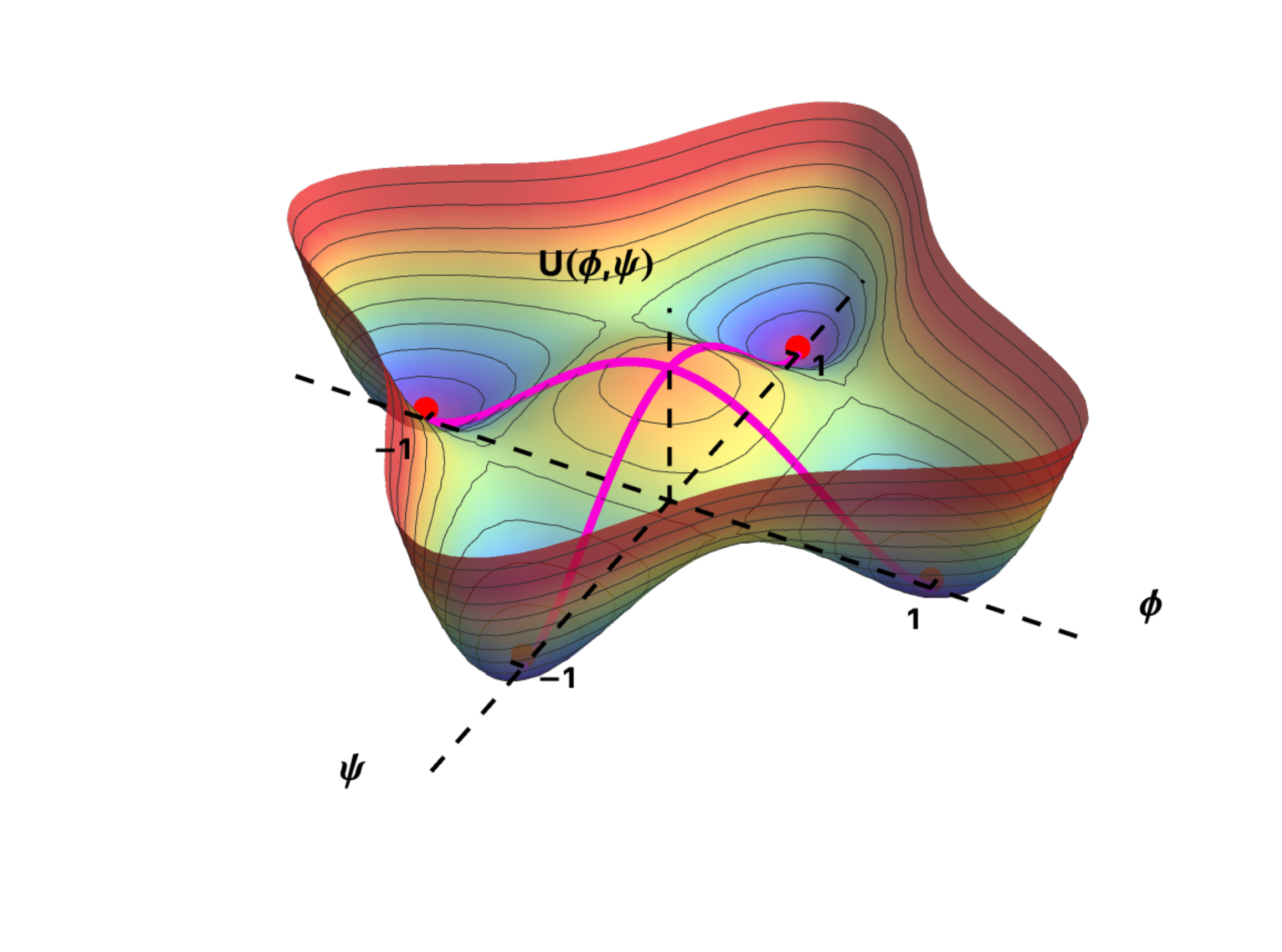}
         \includegraphics[width=0.45\textwidth]{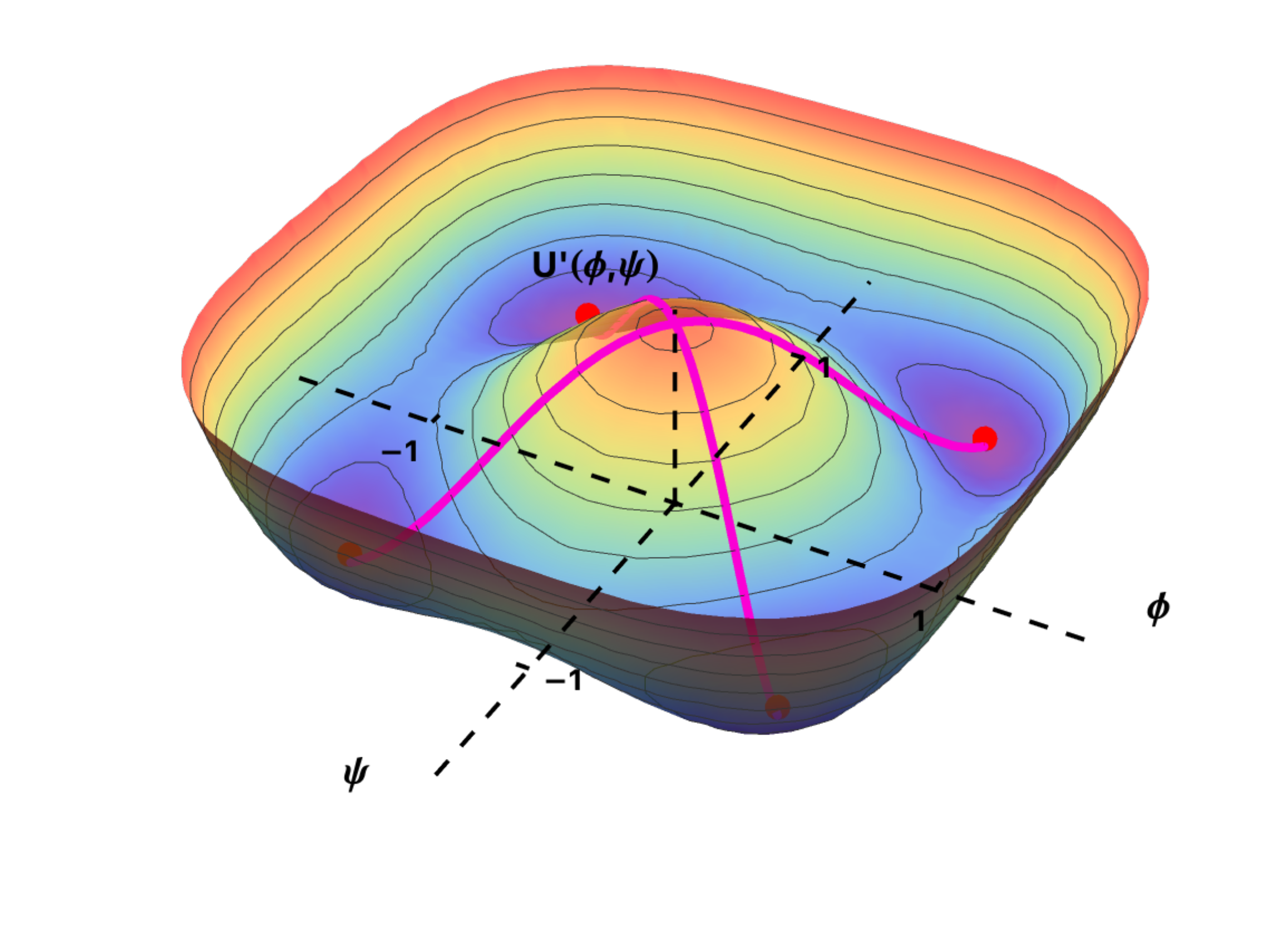}
\caption{Plot of the potential $U(\phi,\psi)$ given by \eqref{Potential} for $\kappa=4$ (upper plot) and for the shifted potential $U'(\phi,\psi)=U(\phi,\psi)+\frac{1-\kappa}{2(1+\kappa)}$ for $\kappa=0.5$ (bottom plot). The corresponding kinks (pink curves) and the minima (red dots) are shown in each case.}
\label{Fig:Potential}
\end{figure}

To clarify as much as possible the problem we are dealing with, in Figure~\ref{Fig:Potential} we show  graphs of the potential for two values of $\kappa$. 
On the first graph of the figure the four vacua \eqref{vacua4>1} of the potentials $U(\phi,\psi)$ in \eqref{Potential} are shown as the minima (red dots) of the potential for $\kappa>1$, and the kinks \eqref{Kink1}--\eqref{Kink2} are  the pink curves connecting two of the non-adjacent vacua.
On the second graph there is a plot of the situation that occurs when $\kappa<1$, with the potential $U'(\phi,\psi)$ appropriately shifted to have the four vacua \eqref{vacua4<1} (red dots), showing again the kinks by pink lines joining two of the non-contiguous vacua.

\subsection{Linear stability and shape modes when $\kappa>1$}

Since  in this paper we will focus on the analysis of the regime $\kappa>1$, we will now discuss in detail the linear stability of the kinks \eqref{Kink1}--\eqref{Kink2}, as well as the shape modes and the corresponding vibrational frequencies that arise in this kind of study. 
First of all, to perform the linear stability analysis, a perturbation in the kink solution $K(x)$ of the form
\begin{equation}\label{complexperturbation}
    \widetilde{K}(x,t)=K(x)+ a \, e^{i\omega t} F(x)= \begin{pmatrix}
\phi(x)  \\ 
\psi(x)
    \end{pmatrix}
+ a \, e^{i \omega t} \begin{pmatrix}
\overline{\eta} (x)  \\ 
\widehat{\eta} (x)
    \end{pmatrix}
\end{equation}
is inserted into the field equations \eqref{FieldEqn1}--\eqref{FieldEqn2}, where $\phi(x)$ and $\psi(x)$ are the two components of the static solution  of the model field equations and  $a$ is a small real parameter. This leads to the spectral problem 
\begin{equation}
    \mathcal{H}
\begin{pmatrix}
\overline{\eta} (x)  \\ 
\widehat{\eta} (x)
    \end{pmatrix}    = 
    \omega^2 
\begin{pmatrix}
\overline{\eta} (x)  \\ 
\widehat{\eta} (x)
    \end{pmatrix}    ,
\end{equation}
where
\begin{equation}
    \mathcal{H}=
    \begin{pmatrix}
          -\dfrac{d^2}{d x^2}+ 6 \phi(x)^2 -2 +2 \kappa\, \psi(x)^2 &  4 \kappa\, \phi(x) \psi(x)  \\
           \hskip-2cm    4 \kappa \, \phi(x) \psi(x)  &  \hskip-2cm -\dfrac{d^2}{d x^2}+ 6 \psi(x)^2 -2 +2 \kappa \,\phi(x)^2  \\
    \end{pmatrix}
    .
\end{equation}
Therefore, to study the kink stability of the solution \eqref{Kink1} in the regime $\kappa>1$, the spectral problem to be solved is essentially the following\footnote{ In fact, all analytical calculations can also be performed with the kink solution \eqref{Kink2}, but all the results would be the same as those corresponding to the kink \eqref{Kink1}.}:
\begin{equation}\label{SpectralProblem}
      \begin{pmatrix}
          -\dfrac{d^2}{d x^2}+ 6 \tanh^2 x -2  & 0  \\
               \hskip-1.7cm 0 &   \hskip-1.7cm -\dfrac{d^2}{d x^2}  +2 \kappa \tanh^2 x -2 \\
    \end{pmatrix}
    \begin{pmatrix}
    \overline{\eta}(x) \\
    \widehat{\eta}(x) 
    \end{pmatrix}
    =
    \omega^2 
    \begin{pmatrix}
    \overline{\eta}(x) \\
    \widehat{\eta}(x) 
    \end{pmatrix}
    .
\end{equation}
From \eqref{SpectralProblem} it can be  seen that the matrix-operator $\mathcal{H}$ is diagonal which, in turn, implies that this problem consists of two decoupled Schr\"odinger-like  equations with P\"oschl-Teller potential wells $\mathcal{H}_{11}=-\frac{d^2}{d x^2}+ 6 \tanh^2 x -2 $ and $\mathcal{H}_{22}=-\frac{d^2}{d x^2}  +2 \kappa \tanh^2 x -2 $. The solution to this type of equations has been widely studied and can be found, for example, in \cite{Flugge1971,Morse1953,Morse1933}. 
Since the orbit of the kink solution \eqref{Kink1} lies  on the $\phi$-axis, the fluctuations corresponding to the first  component of the field will be called \textit{longitudinal eigenmodes} and those corresponding to the second component of the field will be called \textit{orthogonal eigenmodes}. Solving the spectral problem \eqref{SpectralProblem}, the following eigenmodes and eigenfrequencies can be found.
\begin{itemize}
\item 
\textbf{Longitudinal eigenmodes.} 
    There are two eigenmodes, one associated with $\omega=0$, called  \textit{zero mode} or \textit{translational eigenmode}, and another associated with a frequency $\overline{\omega}=\sqrt{3}$, which is known  as \textit{longitudinal shape mode}. 
The formulas for these two modes can be expressed as follows \cite{Vachaspati2006,Shnir2018,Manton1997,Barashenkov2009,Barashenkov2019,Springer2019,AlonsoIzquierdo2023}:
    \begin{eqnarray}
        \overline{F}_{0} (x) \!\!&\!=\!&\!\!         \begin{pmatrix}
    \overline{\eta}_0 (x) \\
    0
    \end{pmatrix}
= \begin{pmatrix}
    \sech^2 x \\
    0
    \end{pmatrix},
    \\ 
        \overline{F}_{\!\!\sqrt{3}}(x)  \!\!&\!=\!&\!\!    \begin{pmatrix}
    \overline{\eta}_D(x) \\
    0
    \end{pmatrix} 
=  \begin{pmatrix}
    \sech x\tanh x \\
    0
    \end{pmatrix} ,
    \label{LongitudinalShapeMode}
    \end{eqnarray}
where the subindex in $F(x)$ indicates the corresponding frequency.
    In  addition to these   vibration eigenfunctions, there are also continuous  modes associated with the frequencies
        \begin{equation}\label{Continuousfrequencylongitudinal}
\overline{\omega}^c_{\bar{q}}=\sqrt{4+\bar{q}^2}, \quad \bar{q}\in\mathbb{R}.
    \end{equation}
In this case, these eigenmodes take the form 
    \begin{equation}\label{ContinuousEigenfunctionFirstField}
        \overline{F}_{\sqrt{4+\bar{q}^2}} \,(x)=
        \begin{pmatrix}
    \overline{\eta}_{\bar{q}}(x) \\
    0
    \end{pmatrix} ,
    \end{equation}
    with $ \overline{\eta}_{\bar{q}}(x)  = ( -1-\bar{q}^2+ 3 \, \tanh^2 x -3 i \bar{q} \tanh x) \, e^{i \bar{q} x}$.
It is easy to verify that the complex conjugate of the previous function is also a solution of \eqref{SpectralProblem}, independent of the first one, which is also obtained with the change $ \bar{q}\to - \bar{q}$. Therefore, this second solution will be denoted as $\overline{\eta}_{-\bar{q}}$ and it can be verified that the Wronskian associated to these two functions is        
\begin{equation}\label{WronskianoFirstField}
        \overline{W}_{\bar{q}}=\overline{\eta}_{\bar{q}}(x)\, \overline{\eta}\,'_{-\bar{q}}(x)-\overline{\eta}\,'_{\bar{q}}(x)\, \overline{\eta}_{-\bar{q}}(x)=-2 i \bar{q}(\bar{q}^2+1)(\bar{q}^2+4),
\end{equation}
where the prime in \eqref{WronskianoFirstField} stands for the derivative with respect to the variable $x$.

As a final remark in this section, let us note that from \eqref{Continuousfrequencylongitudinal} it follows that the continuous spectrum in frequencies begins for $\bar{q}=0$, $\overline{\omega}^c_{0}=2$, that is to say, that the longitudinal discrete spectrum will necessarily be contained in the interval $[0,2 ]$.

\item 
\textbf{Orthogonal eigenmodes.} 
The number of possible shape modes depends on the value of the parameter $\kappa$ that appears in the differential equation corresponding to the second component of \eqref{SpectralProblem}, where the differential operator $\mathcal{H}_{22}$ appears. More precisely, there exists a number $n_{max}\geq 0$ that is given by the largest integer that verifies the following inequality \cite{Flugge1971}
    \begin{equation}\label{ConditionNumberOrthogonalShapeModes}
             \kappa>{n_{max}(n_{max}+1)}/{2},
    \end{equation}
    and which determines the number of discrete eigenmodes ($n_{max}+1$) corresponding to this P\"oschl-Teller type equation.
The eigenfrequencies corresponding to these modes are determined by the expression
\begin{equation}\label{FrequencyOrthogonalModes}
    \widehat{\omega}_n=\sqrt{(2n+1)\rho-n^2-n-\tfrac{5}{2}} , \ n=0,1,\dots, n_{max},
\end{equation}
where $\rho=\sqrt{2\kappa+\frac{1}{4}}$. 
 Notice that the above relationship is telling us something extremely important: that the kink is unstable when $\kappa<3$, since in that case it happens that $\widehat{\omega}^2_0<0$. Therefore, in the following we will focus exclusively on values of $\kappa>3$.

The associated eigenfunctions to these discrete modes \eqref{FrequencyOrthogonalModes} are \cite{Flugge1971}
    \begin{equation}\label{OrthogonalSahpeMode0}
            \widehat{F}_{\widehat{\omega}_n}  (x)=
            \begin{pmatrix}
    0 \\
    \widehat{\eta}_{D,n}(x)
    \end{pmatrix} 
    , \quad n=0,1,\dots, n_{max},
    \end{equation}
    with
    \begin{eqnarray*} \hskip-0.5cm
     \widehat{\eta}_{D,n}(x) \!\!&\!\!=\!\!&\!\!  (\sech x)^{\,\rho-n-\frac{1}{2}}\\
      \!\!&\!\! \!\!&\!\!  \hskip-0.5cm \times\  {}_2F_1  \Bigl( -n,2\rho-n;\rho-n+1/2;\frac{1-\tanh{x}}2 \Bigr),
    \end{eqnarray*}
 being ${}_2F_1 (a,b;c;z)$ the well known hypergeometric function.
Using the following relation (see \cite{NIST2010})
\begin{eqnarray}
& &  \hskip-1.2cm  {}_2F_1 \Bigl(-n,n+2\lambda,\lambda+1/2, \frac{1-x}2 \Bigr)  = \frac{n!}{(2 \lambda)_n} 
        \nonumber   \\
   &\!\! \!\!&\quad \times
     \sum_{m=0}^{\lfloor n/2 \rfloor }(-1)^m \frac{(\lambda)_{n-m}}{m!(n-2m)!}(2x)^{n-2m},
     \label{HypergeometricRelation}
\end{eqnarray}
where $(\lambda)_n= \Gamma(\lambda+n)/\Gamma(\lambda)$ represents the Pochhammer symbol and $ \lfloor x \rfloor$ denotes the integer part of $x$, 
the orthogonal eigenmodes \eqref{OrthogonalSahpeMode0} can be written as
    \begin{eqnarray} \hskip-0.5cm
     \widehat{\eta}_{D,n}(x) \!\!&\!\!=\!\!&\!\! 
    \widehat{\eta}_{D,n}  (x) = (\sech x)^{\,\rho-n-\frac{1}{2}} \frac{n!}{(2\rho-2n)_n} \nonumber
    \\
      \!\!&\!\! \!\!&\!\!  \hskip-0.5cm \times\ \sum_{m=0}^{\lfloor n/2 \rfloor}(-1)^m \frac{(\rho-n)_{n-m}}{m!(n-2m)!}(2\tanh x)^{n-2m}.
    \label{OrthogonalSahpeMode}
\end{eqnarray}
Note that the orthogonal modes $ \widehat{\eta}_{D,n}  (x) $ with even or odd $n$ are even and odd functions, respectively.

On the other hand, the orthogonal continuous spectrum is composed of the following frequencies
\begin{equation}\label{Continuousfrequencyorthogonal}
 \widehat{\omega}^c_{\hat{q}}=\sqrt{\hat{q}^2+2\kappa-2}, \quad \hat{q}\geq 0,
\end{equation}
being the corresponding eigenfunctions
\begin{equation}\label{ContinuousEigenfunctionSecondField}
    \widehat{F}_{\sqrt{\hat{q}^2+2\kappa-2}} \, (x)=
    \begin{pmatrix}
    0 \\
    \widehat{\eta}_{\hat{q}}(x)
    \end{pmatrix} 
\end{equation}
    with
    \begin{eqnarray*} \hskip-0.5cm
     \widehat{\eta}_{\hat{q}}(x) \!\!&\!\!=\!\!&\!\! 
      {}_2F_1  \left(  \frac12-\rho,\frac12+\rho; 1-i \hat{q} ; \frac{1-\tanh x}2  \right) \, e^{i \hat{q} x} .
\end{eqnarray*}
Similar to what happens in the longitudinal case, a second linearly independent solution is obtained by changing $\hat{q}\to - \hat{q}$, so that the Wronskian for these two eigenfunctions can be written in this case as
\begin{equation}\label{WronskianoSecondField}
    \widehat{W}_{\hat{q}}=\widehat{\eta}_{\hat{q}}(x)\, \widehat{\eta}\,'_{-\hat{q}}(x)-\widehat{\eta}\,'_{\hat{q}}(x)\, \widehat{\eta}_{-\hat{q}}(x)=-2 i \hat{q}.
\end{equation}

Note that from \eqref{Continuousfrequencyorthogonal} it follows that the continuous frequency spectrum begins for $\hat{q}=0$, that is, given a value of $\kappa>3$, the discrete orthogonal spectrum will necessarily be contained in the interval $[0,\sqrt{2\kappa-2} ]$.

\end{itemize}

\section{Interaction between vibrational modes: Perturbative approach}\label{Section3}

It has been shown  in some nonlinear models that when we initially excite a shape mode, this  mode couples with the rest of the discrete eigenmodes of the system,  causing  the emission of  radiation  \cite{AlonsoIzquierdo2023}. 
The reason of this phenomenon is the nonlinearity associated with the generalised Klein-Gordon equations  that dictate the behavior of the physical system. 
An example of this can be found  in the $\phi^4$ model, in which, when the shape mode associated with this theory is initially triggered, radiation is found with twice the frequency of the vibrational mode. 

As we mentioned before, in the present work we are going to study how the kink \eqref{Kink1} evolves when an orthogonal shape mode is initially excited. 
The situation in which only the non-zero frequency longitudinal eigenmode is excited reduces the analysis to the study of  a one-component $\phi^4$ model, and this has been extensively investigated by several authors \cite{Manton1997, Barashenkov2009, Barashenkov2019}. 
In light of all this, if the initial configuration takes the form 
\begin{equation}\label{InitialHypothesis}
    \widetilde{K}(x,t)= K(x)+ a_0 \sin(\widehat{\omega}_n t)\, \widehat{F}_{\widehat{\omega}_n}(x),
\end{equation}
where $a_0$ is a small parameter and ${\widehat{\omega}_n}$ is one of the possible values given in \eqref{FrequencyOrthogonalModes}, hopefully at least some of the other modes will be also triggered. Note that \eqref{InitialHypothesis} is similar to \eqref{complexperturbation}, only now we are not looking for complex  but real solutions.

In this section we will address this situation and to do so we will use the perturbative approach that Manton and Merabet first introduced in \cite{Manton1997}. In other words, the following expansion is assumed for field components:
\begin{eqnarray}
\label{MantonExpansion1}
\phi(x,t) \!\!&\!\!=\!\!&\!\! \phi_K(x)+ \overline{a}(t)\,\overline{\eta}_D(x)+\overline{\eta}(x,t),  \\
\psi(x,t) \!\!&\!\!=\!\!&\!\! \sum_p\widehat{a}_p(t)\, \widehat{\eta}_{D, p}(x)+\widehat{\eta}(x,t), \label{MantonExpansion2}
\end{eqnarray}
where $\overline{\eta}_D$, $\widehat{\eta}_{D,p}$ and $\phi_K(x)$ are defined in equations \eqref{LongitudinalShapeMode}, \eqref{OrthogonalSahpeMode} and \eqref{Kink1}. 
The time dependent functions $\overline{a}(t)$ and $\widehat{a}_p(t)$ describe the  evolution of the amplitudes corresponding to each shape mode (longitudinal or transversal), the sum going from $p=0$ to $p=n$, where $n$ is the largest natural number for which the condition \eqref{ConditionNumberOrthogonalShapeModes} is satisfied for a specific value of the coupling constant $\kappa$. 
By construction, $a_0$ is the small parameter in our perturbation approach.  
Furthermore, the functions $\overline{\eta}(x,t)$ and $\widehat{\eta}(x,t)$ are the space and time dependent functions that will dictate the behavior of the radiation found when $x\rightarrow\infty$. For the sake of simplicity,  the dependency of the functions mentioned above will be omitted in subsequent calculations. 
Therefore, if we now plug \eqref{MantonExpansion1}--\eqref{MantonExpansion2} into the field equations     \eqref{FieldEqn1} and \eqref{FieldEqn2}
find for the first and second field components, respectively,
\begin{eqnarray}
    \label{FirstFieldExpansionManton}
   && (\overline{a}_{tt}+\overline{\omega}^2\, \overline{a} \, ) \,\overline{\eta}_D-\overline{\eta}_{xx}+\overline{\eta}_{tt}+2\,\overline{\eta}^3 + 6\, \overline{a}\,\overline{\eta}^2\,\overline{\eta}_D+6\, \overline{a}^2\,\overline{\eta}\,\overline{\eta}_D^2 +2\, \overline{a}^3 \,\overline{{\eta}}_D^3+6\, \overline{\eta}^2\,\phi_K+12\, \overline{a}\,\overline{\eta}\,\overline{\eta}_D\, \phi_K 
    \\ 
   && \quad + 6\,\overline{a}^2\, \overline{\eta}^2_D \,\phi_K + 6\,\overline{\eta}\,\phi_K^2-2\,\overline{\eta}  +2\, \kappa \left(\phi_K+\overline{a}\,\overline{\eta}_D +\overline{\eta} \right) 
   \Bigl(  \sum_{p,r} \widehat{a}_p\,\widehat{a}_r\,\widehat{\eta}_{D ,p}\,\widehat{\eta}_{D,r}+ 2 \sum_p \widehat{a}_p\,\widehat{\eta}_{D, p}\,\widehat{\eta}+\widehat{\eta}^2 \Bigr)=0,
   \nonumber
    \\    
\label{SecondFieldExpansionManton}
       &&  \sum_p \Bigl( (\widehat{a}_{p})_{tt} + \widehat{\omega}_p^2 \, \widehat{a}_p \Bigr) \widehat{\eta}_{D, p}+\widehat{\eta}_{tt}-\widehat{\eta}_{xx}+2\sum_{p,r,s} \widehat{a}_p \,\widehat{a}_r\,\widehat{a}_s\, \widehat{\eta}_{D,p}\,\widehat{\eta}_{D,r}\,\widehat{\eta}_{D,s} + 6 \sum_{p,r} \widehat{a}_p \,\widehat{a}_r\,\widehat{\eta}_{D, p}\,\widehat{\eta}_{D, r} \,\widehat{\eta}  
         \\
         &&
         + 6 \sum_p \widehat{a}_p\, \widehat{\eta}_{D, p}\,\widehat{\eta}^2+2 \widehat{\eta}^3-2\widehat{\eta} + 2\kappa\,\widehat{\eta}\,\phi_K^2   +2 \kappa \Bigl(
         \sum_p \widehat{a}_p\,\widehat{\eta}_{D,p}+\widehat{\eta} \Bigr)   \Bigl( \overline{\eta}^2+ 2 \overline{a}\,\overline{\eta}_D\, \overline{\eta}+ \overline{a}^2 \,\overline{\eta}_D^2+ 2 \overline{\eta}\,\phi_K+2 \overline{a}\,\overline{\eta}_D\,\phi_K \Bigr)  =0,
         \nonumber
    \end{eqnarray}
where $\overline{\omega}=\sqrt{3}$ and $\widehat{\omega}_p$ is given by  \eqref{FrequencyOrthogonalModes}. 
By physical reasons, all the functions $\eta(x,t)$ and $a(t)$ (with a bar or with a hat on top) are small quantities. 
Then, the terms $\eta^2$, $\eta^3$, $\eta^2 a$, $\,a^3$\dots  can be neglected in the formulas \eqref{FirstFieldExpansionManton} and \eqref{SecondFieldExpansionManton}, which leads us to the following truncated expansion for the field component equations:
   \begin{eqnarray}
&&        (\overline{a}_{tt}+\overline{\omega}^2 \,\overline{a} )\,\overline{\eta}_D +\overline{\eta}_{tt}-\overline{\eta}_{xx}-2\overline{\eta}+6\,\overline{\eta}\,\phi_K^2+6\overline{a}^2 \,\overline{\eta}_D^2\,\phi_K+2\kappa\,\phi_K \Bigl( \sum_p \widehat{a}_p \widehat{\eta}_{D, p} \Bigr)^2\approx 0,
        \label{FirstFieldExpansionMantonTruncated}
        \\  
&&        \sum_p \Bigl(  (\widehat{a}_{p})_{tt} + \widehat{\omega}_p^2\,  \widehat{a}_p \Bigr)\, \widehat{\eta}_{D, p}+\widehat{\eta}_{tt}-\widehat{\eta}_{xx}-2 \widehat{\eta}+2\kappa\,\widehat{\eta}       \,\phi_K^2+4 \kappa \,\overline{a}\,\overline{\eta}_D\,\phi_K \sum_p \widehat{a}_p\,
        \widehat{\eta}_{D, p}\approx 0.
        \label{SecondFieldExpansionMantonTruncated}
    \end{eqnarray}
If we now project the formula \eqref{FirstFieldExpansionMantonTruncated} onto the longitudinal shape mode $\overline{\eta}_D(x)$, we find the following relation:
    \begin{equation}
             \Bigl( \overline{a}_{tt}+\overline{\omega}^2 \, \overline{a}\Bigr) \overline{C}+ 6 \, \overline{a}^2 \, \overline{V}+ \sum_{p,r} \widehat{a}_p\, \widehat{a}_r\, \widehat{B}_{p r}=0,
             \label{ODEAmplitudesFirstComponent}
    \end{equation}
    where $ \overline{C} = {2}/{3}$, $\overline{V}= {\pi}/{16}$ and 
    \begin{equation}\label{Bjm}
        \widehat{B}_{p r}= 2 \kappa \int^{\infty}_{-\infty} \overline{\eta}_D(x)\ \phi_K(x)\ \widehat{\eta}_{D, p}(x)\ \widehat{\eta}_{D, r}(x)\ dx.
    \end{equation}
Note that given the parities of the functions in the integrand of \eqref{Bjm}, if $p+r$ is an odd number, then $\widehat{B}_{p r}=0$.
    
Similarly, we can project the relation \eqref{SecondFieldExpansionMantonTruncated} onto $\widehat{\eta}_{D, m}$, which leads to
\begin{equation}
    \Bigl(   (\widehat{a}_{m})_{tt}   + \widehat{\omega}_m^2 \,  \widehat{a}_{m} \Bigr) \widehat{C}_{m} +2 \sum_{p} \overline{a}\, \widehat{a}_p \,\widehat{B}_{p m}=0,
    \label{ODEAmplitudesSecondComponent}
\end{equation}
$m=0,1, \dots n,$, where 
\begin{equation}\label{CDj}
 \widehat{C}_{m} =\int^{\infty}_{-\infty} \widehat{\eta}_{D, m}^2(x)\ dx.
\end{equation} 
When necessary, the numbers $\widehat{B}_{p r}$ in \eqref{Bjm} and $\widehat{C}_{ m}$ in \eqref{CDj} should be evaluated numerically, since the presence of the functions $\widehat{\eta}_{D,n}$ in the corresponding integrals makes it impossible to evaluate them analytically.

The expressions \eqref{ODEAmplitudesFirstComponent} and \eqref{ODEAmplitudesSecondComponent} form a system of $n+1$ ordinary nonlinear differential equations that must be determined according to the shape modes that are initially triggered (the initial conditions). Plugging \eqref{ODEAmplitudesFirstComponent} and \eqref{ODEAmplitudesSecondComponent} into \eqref{FirstFieldExpansionMantonTruncated} and \eqref{SecondFieldExpansionMantonTruncated},  we find 
\begin{eqnarray}\label{EquationRadiationFirstField}
         \overline{\eta}_{tt}-\overline{\eta}_{xx}+
         ( 6 \phi_K^2-2 ) \overline{\eta}  \!\!&\!\!=\!\!&\!\!  \overline{a}^2 \Bigl( -6\,\overline{\eta}_D^2\,\phi_K+\frac{\displaystyle6\,  \overline{V}}{\displaystyle\overline{C}}\,\overline{\eta}_D \Bigr)+\sum_{p,r} \widehat{a}_p\, \widehat{a}_r\,\Bigl(\displaystyle \frac{\widehat{B}_{p r}}{\displaystyle\overline{C}_{D}}\,\overline{\eta}_D-2\,\kappa\, \phi_K \,\widehat{\eta}_{D, p}\,\widehat{\eta}_{D, r}\Bigr)
\\ 
\label{EquationRadiationSecondField}
             \widehat{\eta}_{tt}-\widehat{\eta}_{xx}+  ( 2\kappa\, \phi_K^2-2  ) \widehat{\eta}
             \!\!&\!\!=\!\!&\!\! 2\,\overline{a} \, \Bigl(\sum_{p,r}\frac{\displaystyle  \widehat{a}_r\, \widehat{B}_{pr}\,\widehat{\eta}_{D, r}}{\displaystyle\widehat{C}_{p}}-2 \kappa \sum_p\,\overline{\eta}_D \,\phi_K \,\widehat{a}_p \,\widehat{\eta}_{D, p}\Bigr)
\end{eqnarray}
for the first and second components. 
These will be the key equations that we need to analyze next.

\subsection{Evolution of the system when only an orthogonal shape mode is initially activated}

Next, we want to study how the systems evolves in case we exclusively trigger the $j$-th orthogonal shape mode at $t=0$. It can be seen from \eqref{ODEAmplitudesFirstComponent}--\eqref{ODEAmplitudesSecondComponent} that this excitation also activates the rest of the shape modes, but their corresponding amplitudes will be much smaller. From this reasoning and from the differential equation \eqref{ODEAmplitudesSecondComponent}, it is logical to assume that the amplitude associated with the $j$-th mode can be approximated as
\begin{equation}\label{OthogonalAmplitudeEvolution}
    \widehat{a}_j(t)\approx a_0 \sin (\widehat{\omega}_j t),
\end{equation}
where ${\widehat{\omega}_j}$ is fixed and is one of the possible values given in \eqref{FrequencyOrthogonalModes}. 
If we now plug \eqref{OthogonalAmplitudeEvolution} into \eqref{ODEAmplitudesFirstComponent}, we neglect terms of the form $\overline{a}^2$ and $\widehat{a}_p\, \widehat{a}_r$ with $ p ,r\neq j$ (because they are of order $\mathcal{O}(a_0^4)$) and then we solve the resulting differential equation taking into account the initial conditions
\begin{equation}\label{IntialConditions}
    \overline{a}_t(0)=\overline{a}(0)=0, \  \widehat{a}_m(0)=\widehat{a}_m(0)_t=0 \  {\rm with} \  m\neq j, 
\end{equation}
then it turns out that the evolution of $\overline{a}(t)$ can be described as
\begin{equation}\label{LongitudinalAmplitudeEvolution}
        \overline{a}(t)\approx\frac{a_0^2\, \widehat{B}_{jj}\left( 4 \widehat{\omega}_j^2 -\overline{\omega}^2+\overline{\omega}^2 \cos (2 \widehat{\omega}_j t)-4\widehat{\omega}_j^2 \cos(\overline{\omega}t)\right)}{2\overline{C}\,  \overline{\omega}^2 \left( \overline{\omega}^2-4 \widehat{\omega}_j^2\right)}.
\end{equation}
The initial conditions \eqref{IntialConditions} have been taken because only the $j$-th mode is activated and none of the others.

On the other hand, the rest of the amplitudes $\widehat{a}_m$ can be estimated by solving the differential equations \eqref{ODEAmplitudesSecondComponent}, neglecting in them the terms $\overline{a}\; \widehat{a }_p$, with $p\neq j$, with the initial conditions \eqref{IntialConditions}, which leads to 
\begin{equation}\label{OrthogonalAmplitudes}
\widehat{a}_m(t) = \widehat{a}_{\widehat{\omega}_j} \sin(\widehat{\omega}_j t)+ \widehat{a}_{3\widehat{\omega}_j} \sin(3\widehat{\omega}_j t)+ \widehat{a}_{\widehat{\omega}_m} \sin(\widehat{\omega}_m t)+ \widehat{a}_{\widehat{\omega}_j+\overline{\omega}} \sin((\widehat{\omega}_j+\overline{\omega})t)+ \widehat{a}_{\widehat{\omega}_j-\overline{\omega}} \sin((\widehat{\omega}_j-\overline{\omega})t),
\end{equation}
where the amplitudes associated with each of the frequencies that appear in the expression \eqref{OrthogonalAmplitudes} are
\begin{eqnarray}
\hskip-0.7cm       \widehat{a}_{\widehat{\omega}_j} \!\!&\!\!=\!\!&\!\! \frac{a_0^3 \, \widehat{B}_{jm} \, \widehat{B}_{jj} \left( 3 \overline{\omega}^2-8 \widehat{\omega}_j^2\right)}{2\overline{C} \, \widehat{C}_m \left( \widehat{\omega}_j^2-\widehat{\omega}_m^2\right) \overline{\omega}^2 \left( 4\widehat{\omega}_j^2-\overline{\omega}^2\right)}, \label{OrthogonalAmplitudes1}
    \qquad\qquad 
    \widehat{a}_{3\widehat{\omega}_j}  =  \frac{-a_0^3 \,  \widehat{B}_{jm} \, \widehat{B}_{jj}}{2\overline{C} \, \widehat{C}_m \left( 9\widehat{\omega}_j^2-\widehat{\omega}_m^2 \right) \left( 4\widehat{\omega}_j^2-\overline{\omega}^2 \right)},
    \\
\hskip-0.7cm       \widehat{a}_{\widehat{\omega}_m}  \!\!&\!\!=\!\!&\!\! \frac{- 4 a_0^3 \,  \widehat{B}_{jm} \, \widehat{B}_{jj}  \, \widehat{\omega}_j^3 \left( 3\widehat{\omega}_j^2+5\widehat{\omega}_j^2-3\overline{\omega}^2 \right)}{\overline{C} \, \widehat{C}_m  \, \widehat{\omega}_m \left( 9\widehat{\omega}_j^4-10\widehat{\omega}_j^2 \, \widehat{\omega}_m^2+\widehat{\omega}_m^4 \right) \left( \widehat{\omega}_j^4+\left( \widehat{\omega}_m^2-\overline{\omega}^2 \right)^2 -2\widehat{\omega}_j^2 \left( \widehat{\omega}_m^2+\overline{\omega}^2 \right) \right) }, 
    \label{OrthogonalAmplitudes3}
    \\
\hskip-0.7cm       \widehat{a}_{\widehat{\omega}_j+\overline{\omega}}  \!\!&\!=\!\!&\!\! \frac{-2 a_0^3 \, \widehat{B}_{jm} \, \widehat{B}_{jj}  \, \widehat{\omega}_j^2 }{\overline{C} \, \widehat{C}_m  \, \overline{\omega}^2  ( \widehat{\omega}_j-\widehat{\omega}_m+\overline{\omega}  )  ( \widehat{\omega}_j+\widehat{\omega}_m+\overline{\omega}  )  ( \overline{\omega}^2-4\widehat{\omega}_j^2  )}, 
    \label{OrthogonalAmplitudes4}
    \
    \widehat{a}_{\widehat{\omega}_j-\overline{\omega}}  =\frac{-2 a_0^3 \, \widehat{B}_{jm} \, \widehat{B}_{jj} \, \widehat{\omega}_j^2 }{\overline{C}  \widehat{C}_m  \overline{\omega}^2  ( \overline{\omega}^2-4\widehat{\omega}_j^2  )  ( \widehat{\omega}_j^2-\widehat{\omega}_m^2 -2\widehat{\omega}_j  \overline{\omega}+\overline{\omega}^2  )}.
    \label{OrthogonalAmplitudes5}
\end{eqnarray}
From these formulas and from the parity of the shape modes, it can be shown that $\widehat{a}_m$ is zero when we consider shape modes with different parities. 
This is because  the integrand of \eqref{Bjm} is odd  when $j+m$ is not an even number. This phenomenon will be studied in detail in Section~\ref{Section4.4}, where we will compare these results with those obtained by numerical simulations.

When we substitute \eqref{OrthogonalAmplitudes1}--\eqref{OrthogonalAmplitudes5} into \eqref{OrthogonalAmplitudes}, the resulting amplitudes lead to terms of order $\mathcal{O}(a_0^4)$ in  \eqref{EquationRadiationFirstField}--\eqref{EquationRadiationSecondField}, which will be ignored because we are only considering quantities up to the order $\mathcal{O}(a_0^3)$. In other words, the differential equations \eqref{EquationRadiationFirstField} and \eqref{EquationRadiationSecondField} can be approximated up to the order indicated by the following ones
\begin{eqnarray}
 \hskip-0.7cm    \overline{\eta}_{tt}-\overline{\eta}_{xx}+  ( 6 \phi_K^2-2 )\overline{\eta} \!\!&\!\! =\!\!&\!\!  \widehat{a}_j^2 \left[ \frac{\displaystyle \widehat{B}_{jj}}{\displaystyle \overline{C}_{D}}\overline{\eta}_D -
    2 \kappa  \phi_K  \widehat{\eta}_{D, j}^2\right],
    \label{EquationRadiationFirstField2} 
    \\ 
 \hskip-0.7cm        \widehat{\eta}_{tt}-\widehat{\eta}_{xx}+  ( 2 \kappa  \phi_K^2-2  ) \widehat{\eta}  \!\!&\! =\!&\!\! 2 \overline{a} \widehat{a}_j \left[\frac{  \widehat{B}_{jj}}{\widehat{C}_{ j}}  \widehat{\eta}_{D, j}-
    2  \kappa  \overline{\eta}_D  \phi_K    \widehat{\eta}_{D, j}\right]. \nonumber\\ \label{EquationRadiationSecondField2}
\end{eqnarray}
Taking into account that from \eqref{OthogonalAmplitudeEvolution}
\begin{equation}\label{shapeModeAmplitudeSquare}
     \widehat{a}_j^2(t)=\frac{a_0^2}{2} \left( 1-\cos(2 \widehat{\omega}_j t )\right),
\end{equation}
and that the time-independent part of  \eqref{shapeModeAmplitudeSquare} causes a time-independent response of  $\overline{\eta}$ that carries zero energy, then it is possible to rewrite \eqref{EquationRadiationSecondField2} as
\begin{equation}\label{EquationRadiationFirstField3}
     \overline{\eta}_{tt}-\overline{\eta}_{xx}+ \left( 6 \phi_K^2-2 \right)\,\overline{\eta}=f(x)\ e^{i2\widehat{\omega}_j t},
\end{equation}
where
\begin{equation}
    f(x)= -\frac{a_0^2}{2}\left(\frac{\widehat{B}_{jj}}{\overline{C}_{D}}\,\overline{\eta}_D-2\kappa\, \phi_K\, \widehat{\eta}_{D, j}^2\right).
\end{equation}
It is important to clarify that, to simplify the subsequent calculations as much as possible, we have taken imaginary exponentials in \eqref{EquationRadiationFirstField3} instead of sines or cosines. In fact, this does not affect the final analytical result at all since the relevant results can be retrieved simply by taking the real part of the final result.
On the other hand, from \eqref{OthogonalAmplitudeEvolution} and \eqref{LongitudinalAmplitudeEvolution} it can be obtained that 
\begin{equation}
     \overline{a}(t)\, \widehat{a}_j (t)=\frac{\displaystyle a_0^3 \, \widehat{B}_{jj}\left[4 \widehat{\omega}_j^2 \left( \sin{((\widehat{\omega}_j-\overline{\omega})t)}+\sin{((\widehat{\omega}_j+\overline{\omega})t)} \right) + \left( 3 \overline{\omega}^2-8 \widehat{\omega}_j^2  \right)\sin(\widehat{\omega}_j t )-\overline{\omega}^2 \sin(3\widehat{\omega}_j t) \right]}{\displaystyle 4\overline{C}  \, \overline{\omega}^2 \left(4 \widehat{\omega}_j^2-\overline{\omega}^2 \right)}.
\end{equation}
Clearly, in the expression we have just found, four relevant frequencies naturally appear, which are
\begin{equation}\label{PossibleFrequencies}
    \omega_1=\widehat{\omega}_j,\ \omega_2=3\widehat{\omega}_j,\  \omega_3=\widehat{\omega}_j+\overline{\omega},\   \text{and} \  \omega_4=|\widehat{\omega}_j-\overline{\omega}|.
\end{equation}
Following the same procedure used with the equation of the first field component $\overline{\eta}$, the formula \eqref{EquationRadiationSecondField2} can be rewritten as
\begin{equation}\label{EquationRadiationSecondField3}
     \widehat{\eta}_{tt}-\widehat{\eta}_{xx}+ \left( 2 \kappa \, \phi_K^2-2 \right)\,\widehat{\eta}=\sum_{\ell=1}^4 g_\ell(x) \ e^{i \omega_\ell t},
\end{equation}
where
\begin{eqnarray*}
\hskip-0.7cm  &&    g_1(x)  =  \frac{ a_0^3 \, \widehat{B}_{jj}\left( 3\overline{\omega}^2-8\widehat{\omega}_j^2 \right)}{2\overline{C}\,\overline{\omega}^2\left( 4\widehat{\omega}_j^2-\overline{\omega}^2 \right)}\left(\frac{  \widehat{B}_{jj}}{\widehat{C}_{j}}\,\widehat{\eta}_{D, j}-2 \kappa \,\overline{\eta}_D \,\phi_K \, \widehat{\eta}_{D, j}\right),
     \qquad
      g_2(x)   =  \frac{- a_0^3\, \widehat{B}_{jj}}{2\overline{C} \left(4\widehat{\omega}_j^2-\overline{\omega}^2 \right)}\left(\frac{  \widehat{B}_{jj}}{\widehat{C}_{j}}\,\widehat{\eta}_{D, j}-2 \kappa \,\overline{\eta}_D \, \phi_K \, \widehat{\eta}_{D, j}\right),
      \\
 \hskip-0.7cm  &&     g_3(x)=g_4(x)   =  \frac{2 a_0^3\, \widehat{B}_{jj}\,\widehat{\omega}_j^2}{\overline{C}\,\overline{\omega}^2\left( 4\widehat{\omega}_j^2-\overline{\omega}^2 \right)}\left(\frac{  \widehat{B}_{jj}}{\widehat{C}_{j}}\,\widehat{\eta}_{D,j}-2 \kappa \,\overline{\eta}_D\, \phi_K  \,\widehat{\eta}_{D, j}\right).
\end{eqnarray*}
Under these circumstances, \eqref{EquationRadiationFirstField3} and \eqref{EquationRadiationSecondField3} can be solved if we separate the spatial and temporal part of the two $\eta(x,t)$-functions as follows
\begin{equation}
\overline{\eta}(x,t)= \overline{\eta}_{2\widehat{\omega}_j}(x)\ e^{i\, 2\widehat{\omega}_j t},
\quad
    \widehat{\eta}(x,t)= \sum_{\ell=1}^4\widehat{\eta}_{\omega_\ell} (x)\ e^{i \omega_\ell t},
\end{equation}
which  will lead to the non-homogeneous linear ordinary differential equations ($\ell=1,2,3,4$):
\begin{eqnarray}
 -\overline{\eta}''_{2\widehat{\omega}_j}(x)+ \left( 6  \phi_K^2-2-4\widehat{\omega}_j^2 \right)\,\overline{\eta}_{2\widehat{\omega}_j}(x) \!\!&\!\!=\!\!&\!\!  f(x),
 \label{EquationRadiationFirstField4}
 \\
    -\widehat{\eta}''_{\omega_\ell}(x)+ \left( 2 \kappa \, \phi_K^2-2-\omega_\ell^2 \right) 
    \widehat{\eta}_{\omega_\ell}(x) \!\!&\!\!=\!\!&\!\!  g_\ell(x),
    \label{EquationRadiationSecondField4} 
\end{eqnarray}
If we now take into account the dispersion relations  for a longitudinal channel \eqref{Continuousfrequencylongitudinal} of frequency $ 2\widehat{\omega}_j$ and the four orthogonal modes \eqref{Continuousfrequencyorthogonal}, which are the $\omega_\ell$ given in \eqref{PossibleFrequencies},
\begin{equation}
\bar{q}=\sqrt{4 \widehat{\omega}^2_j-4} , \qquad \widehat{q } _ \ell=\sqrt {\omega^2_\ell+2-2\kappa},
\end{equation}
 as well as the homogeneous solutions of \eqref{EquationRadiationFirstField4} and \eqref{EquationRadiationSecondField4}, which correspond to the expressions \eqref{ContinuousEigenfunctionFirstField} and \eqref{ContinuousEigenfunctionSecondField}, then the  solutions to the inhomogeneous equations \eqref{EquationRadiationFirstField4}--\eqref{EquationRadiationSecondField4} are given by the following functions:
\begin{eqnarray}
    \overline{\eta}_{2 \widehat{\omega}_j} \!\!&\!\!=\!\!&\!\! -\frac{  \overline{\eta}_{-{\bar{q}}}(x)}{\overline{W}_{\bar{q}}}\int_{-\infty}^{x}\overline{\eta}_{{\bar{q}}} (y)\, f(y)\, dy \nonumber \\
    \!\!&\!\!   \!\!&\!\! 
    -\frac{ \overline{\eta}_{{\bar{q}}} (x)}{\overline{W}_{\bar{q}}}\int_{x}^{\infty}\overline{\eta}_{-{\bar{q}}} (y)\, f(y)\, dy,\label{SolutionRadiationFirstField}  \\[1ex]
    \widehat{\eta}_{\omega_\ell}  \!\!&\!\!=\!\!&\!\!  -\frac{ \widehat{\eta}_{-{\widehat{q}_\ell}}(x)}{\widehat{W}_{\widehat{q}_\ell}}\int_{-\infty}^{x}\widehat{\eta}_{{\widehat{q}_\ell}} (y)\, g_\ell(y)\, dy \nonumber \\
    \!\!&\!\!   \!\!&\!\! -\frac{ \widehat{\eta}_{{\widehat{q}_\ell}}(x)}{\widehat{W}_{\widehat{q}_\ell}}\int_{x}^{\infty}\widehat{\eta}_{-{\widehat{q}_\ell}}(y)\, g_\ell(y)\, dy,  \label{SolutionRadiationSecondField}
\end{eqnarray}
where $\overline{W}_{\bar{q}}$ and $\widehat{W}_{\widehat{q}_\ell}$ are given by \eqref{WronskianoFirstField} and \eqref{WronskianoSecondField}. The asymptotic behavior of \eqref{SolutionRadiationFirstField} and \eqref{SolutionRadiationSecondField} is
\begin{eqnarray}
     \overline{\eta}_{2 \widehat{\omega}_j}&\xrightarrow{x\rightarrow\infty}& \frac{i\left(\int_{-\infty}^{\infty }\overline{\eta}_{{\overline{q}}}(y)f(y)dy\right)}{2(\overline{q}+i)(\overline{q}+2i)} e^{-i\overline{q}x}\label{FinalRadiationFirstField},\\ 
    \widehat{\eta}_{\omega_\ell}&\xrightarrow{x\rightarrow\infty}& \frac{\left(\int_{-\infty}^{\infty}\widehat{\eta}_{{\widehat{q}_\ell}}(y)g_\ell(y)dy\right)}{2 i \widehat{q}_\ell }\,  e^{-i \widehat{q}_\ell x},\label{FinalRadiationSecondField}
\end{eqnarray}
which provides us with the amplitudes of the radiation that travels in the longitudinal and orthogonal channels respectively.
Unfortunately, the functions \eqref{FinalRadiationFirstField} and \eqref{FinalRadiationSecondField} cannot be calculated analytically since the integrals that are present in these formulas involve mixed hypergeometric and hyperbolic functions. Note that it can be seen that the amount of radiation propagated in the longitudinal channel is going to be greater than that emitted in the orthogonal channel because the term \eqref{FinalRadiationFirstField} is proportional to $a_0^2$, while the term \eqref{FinalRadiationSecondField} is proportional to $a_0^3$, being  $a_0$ a small parameter.

Now that the possible radiation frequencies are known, we have to find out which of them are capable of producing radiation.
In fact, for this to happen both $2\widehat{\omega}_j$ and $\omega_{\ell}$  in \eqref{PossibleFrequencies} have to lie within the continuous vibration spectra of the components of the first and second fields, respectively \cite{AlonsoIzquierdo2023}.
In other words, $\bar{q}$ and $\widehat{q}_\ell$ must both be positive real quantities. This can be verified from \eqref{FinalRadiationFirstField} and \eqref{FinalRadiationSecondField}, since, if the aforementioned dispersion relations were imaginary, then this would lead to divergences in the solutions \eqref{SolutionRadiationFirstField}--\eqref{SolutionRadiationSecondField} when we are far from the center of the kink.

\begin{figure}[htb]
\centering
 \includegraphics[width=0.4\textwidth]{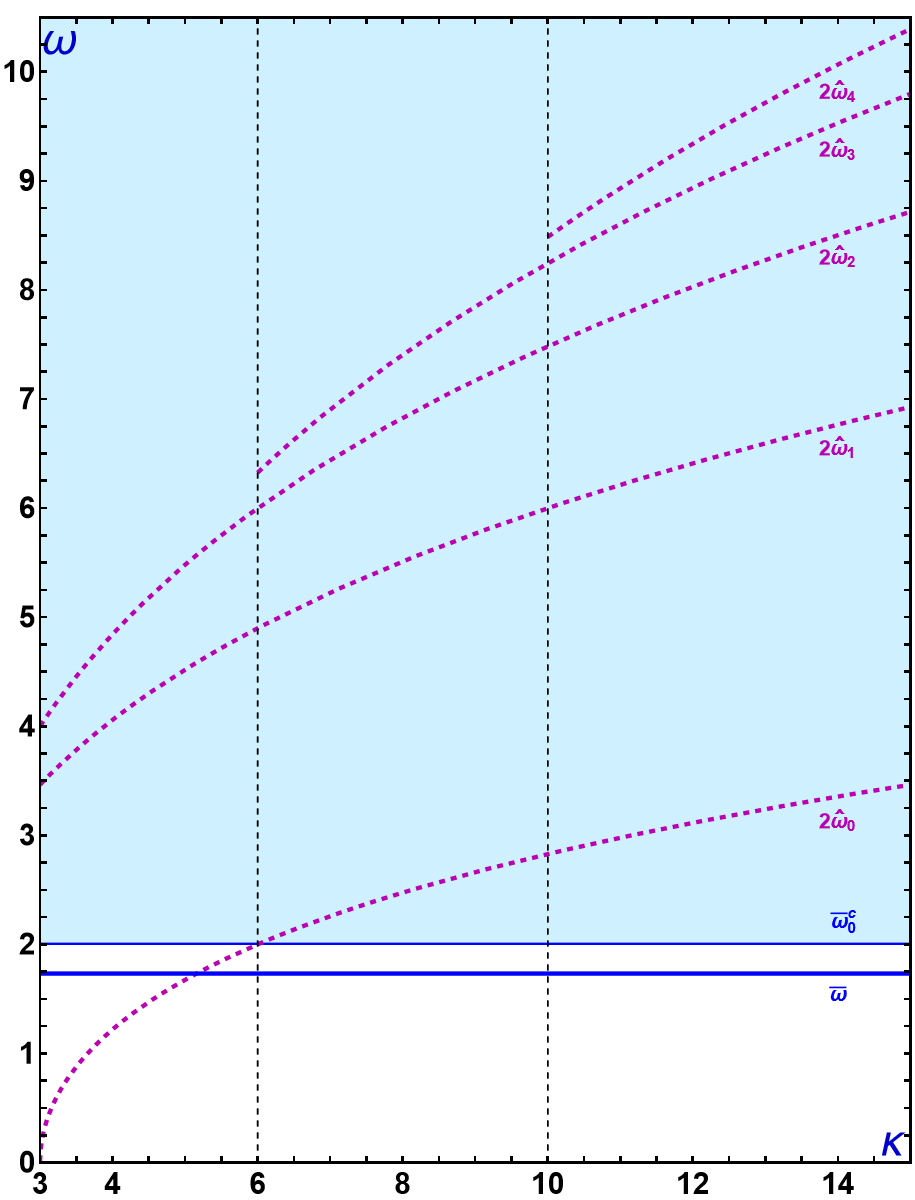}
\caption{Longitudinal eigenfrequencies of the operator $\mathcal{H}_{11}$ in \eqref{SpectralProblem} as a function of the coupling constant $\kappa$. The black dashed lines are the values of $\kappa$ for which a new orthogonal mode arises in the spectrum of $\mathcal{H}_{22}$. The purple lines represent the lowest frequencies that can be excited by the coupling with the orthogonal fluctuations, which can be realized as radiation when plunged into the continuous spectrum (blue area).}
                           \vspace{-0.4cm}
\label{Fig:LongitudinalEspectrum}
\end{figure}
Below we are going to graphically illustrate part of the analytical results that we have obtained so far, to help us better understand the solutions to the problem that we are analyzing. 
 Thus, Figure~\ref{Fig:LongitudinalEspectrum} shows the eigenfrequencies found in Section~\ref{Section2} for the longitudinal fluctuations of $\mathcal{H}_{11}$ in \eqref{SpectralProblem}, from which we can infer that if we initially activate only $\widehat{\eta}_{D,0} $, then we can only find radiation with frequency $2 \widehat{\omega}_0 $ in the regime $\kappa>6 $ for the longitudinal channel, since for $\kappa<6$ it happens that $2\widehat{\omega} _0< \overline{\omega}^c_0$.
 For higher modes, the frequencies $2\widehat{\omega}_n$ are always embedded in the continuous part of the spectrum $\mathcal{H}_{22}$.

\begin{figure}[h!]
         \centering
         \includegraphics[width=0.4\textwidth]{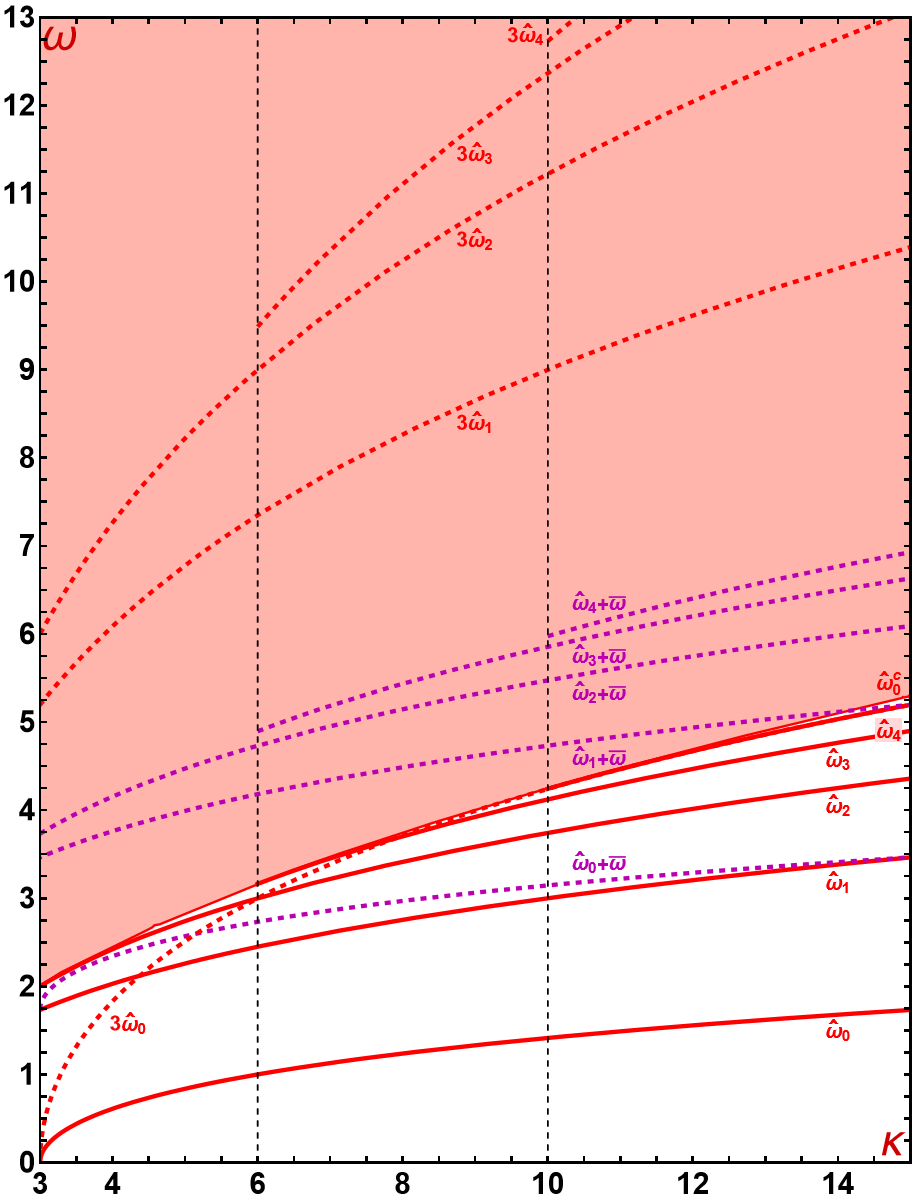}
         \qquad
         \includegraphics[width=0.4\textwidth]{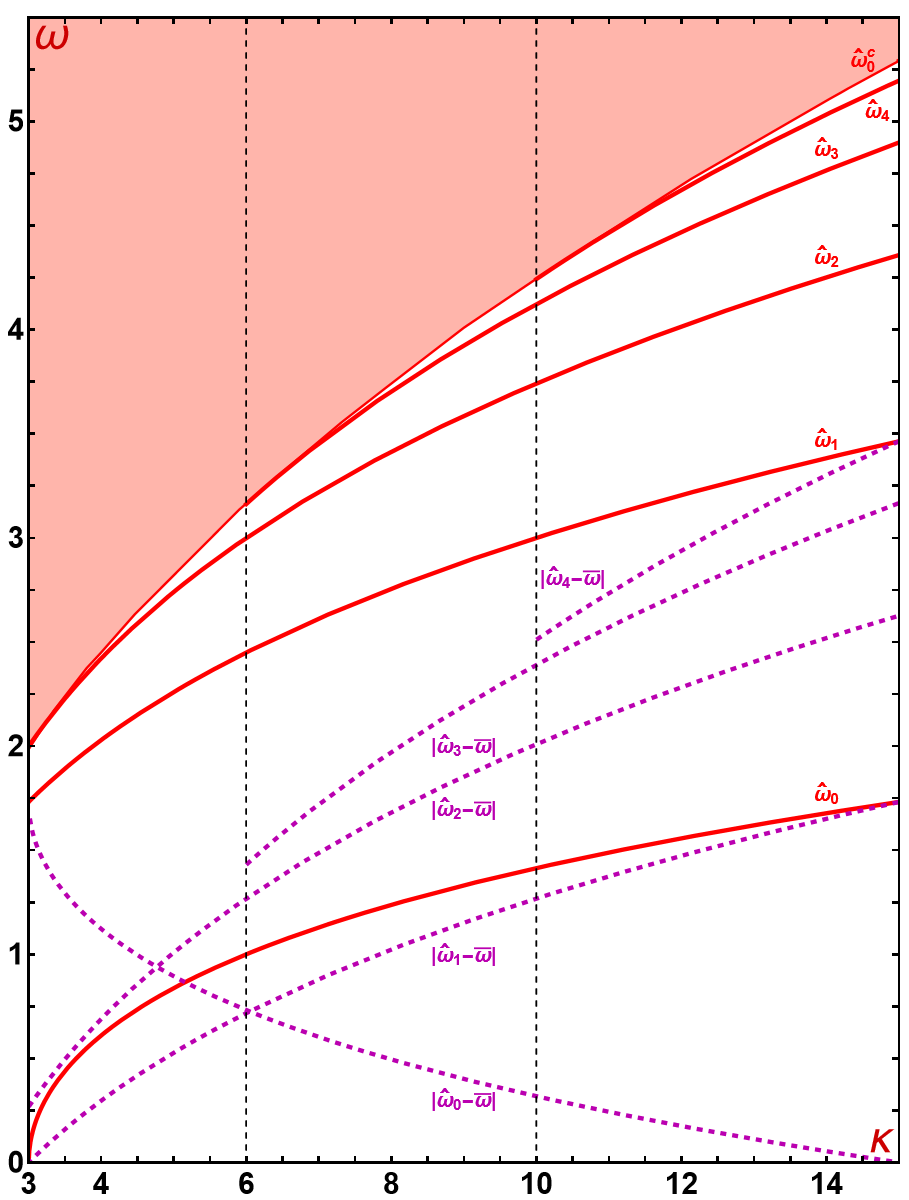}
\caption{On the left is part of the spectrum of the operator $\mathcal{H}_{22}$ that involves the orthogonal fluctuations as a function of the coupling constant $\kappa$. The graph on the right shows a zoom of the aforementioned spectrum near the threshold value $\widehat{\omega}^c_0$.
The black dashed vertical lines are the values of $\kappa$ for which a new orthogonal shape mode arises. 
The red and purple dotted lines represent the lowest frequencies that can be excited by coupling with longitudinal fluctuations, which can be realized as radiation when immersed in the continuous spectrum (red area).}
                           \vspace{-0.5cm}
     \label{Fig:OrthogonalSpectrum}
\end{figure}
The two graphs in Figure~\ref{Fig:OrthogonalSpectrum} show the spectrum of the orthogonal operator $\mathcal{H}_{22}$, in addition to the frequencies $\omega_1,\dots\omega_4$ defined in \eqref{PossibleFrequencies}. 
As we have already said, only the frequencies embedded in the continuous spectrum will be able to produce radiation. Following this reasoning, as we can see in the second of the drawings in Figure~\ref{Fig:OrthogonalSpectrum}, the frequencies $|\widehat{\omega}_i-\overline{\omega}|$ are not embedded in the continuous part of the orthogonal channel spectrum, which implies that no radiation associated with these frequencies will be found. 
On the other hand, from the first of the drawings in Figure~\ref{Fig:OrthogonalSpectrum} it can be inferred that $3\widehat{\omega}_0$ will not produce radiation in the second field component, although this frequency is ``almost" embedded in the continuous part of the  spectrum around $\kappa=10$. 
Also note that $3\widehat{\omega}_0$ coincides with $\widehat{\omega}_4$ for $\kappa>10$.  The rest of the frequencies $3\widehat{\omega}_i$ are part of the radiation spectrum, which implies that radiation associated with these frequencies can be detected. On the other hand,  the frequencies $\widehat{\omega_i}+\overline{\omega}$ (except for $\widehat{\omega_0}+\overline{\omega}$)  are contained in the continuous spectrum only for a range of values of the coupling constant $\kappa$. For example, it can be shown that $\widehat{\omega}_1+\overline{\omega}>\widehat{\omega}^c_0$ when $3<\kappa<14.14$ and $\widehat{\omega}_2+\overline{\omega}>\widehat{\omega}^c_0$ when $3<\kappa<24.93$.

\begin{figure}[htb]
     \centering
         \includegraphics[width=0.45\textwidth]{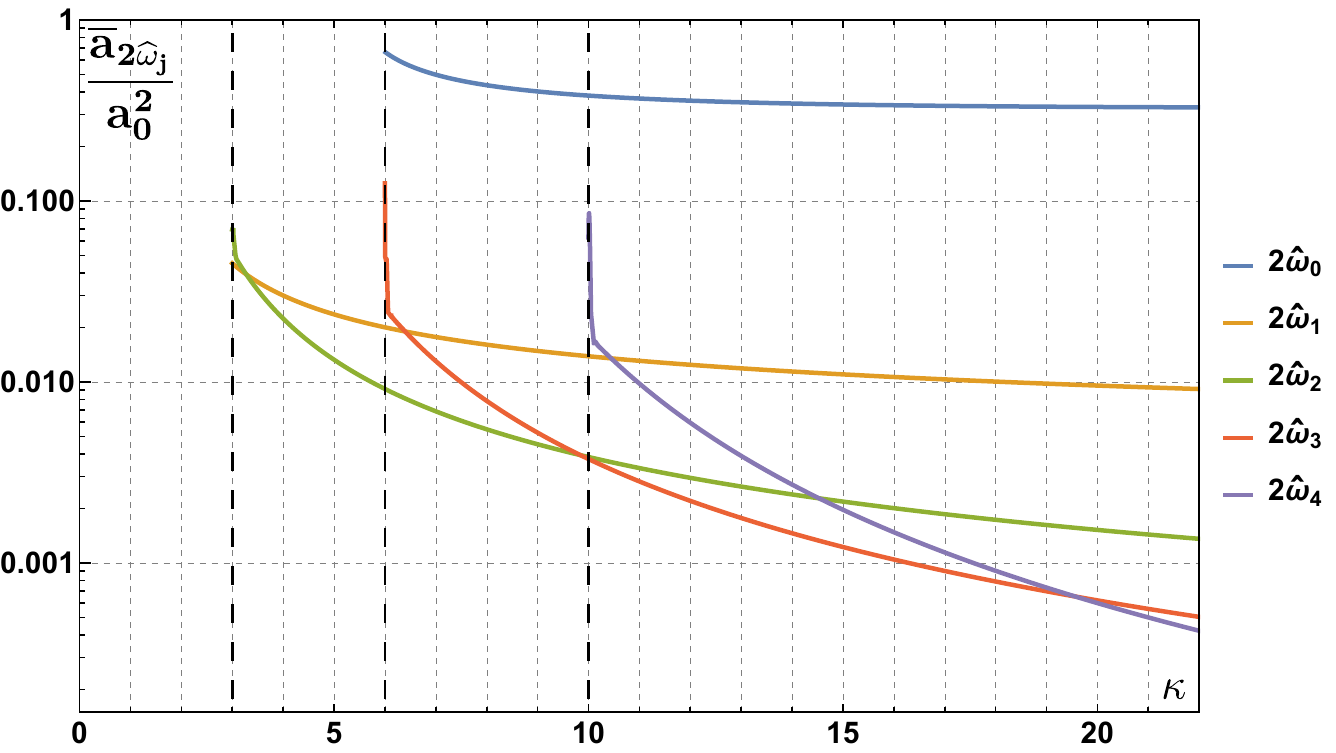}
                  \vspace{-0.3cm}
         \caption{Radiation amplitudes  associated with the frequencies  $ 2\widehat{\omega}_0,\dots 2\widehat{\omega}_4$ in the longitudinal  channel. The vertical dashed lines indicate the values of $\kappa$ for which a new shape mode arises in the spectrum of $\mathcal{H}_{22}$.}
         \label{Fig:LongitudinalAmplitudes}
\end{figure}

\begin{figure}[htb]
     \centering
         \includegraphics[width=0.45\textwidth]{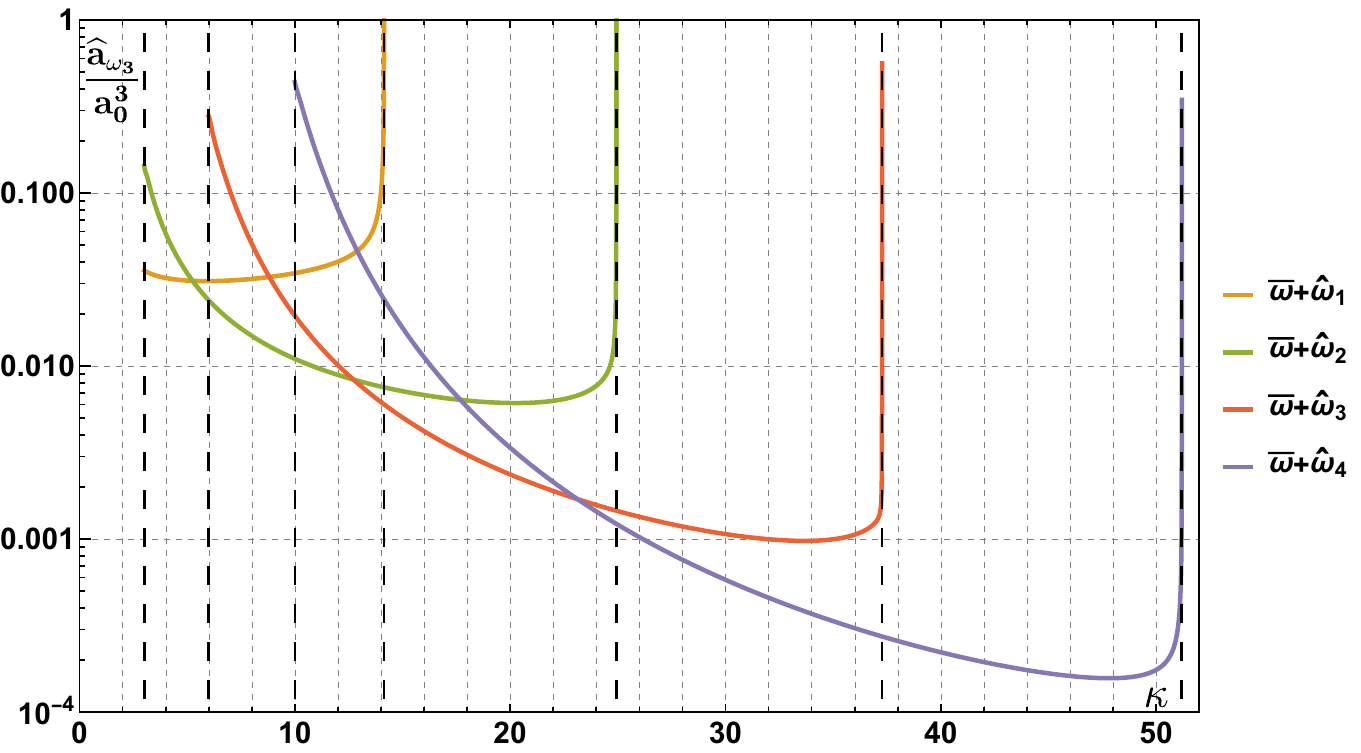}
                  \vspace{-0.3cm}
         \caption{Radiation amplitudes  associated with the frequencies $ \overline{\omega}+\widehat{\omega}_j$ ($j=1,\dots,4$) in the orthogonal channel. The dashed lines indicate the values of $\kappa$ for which a new shape mode arises in the spectrum of $\mathcal{H}_{22}$ and for which a radiation term disappears.}
         \label{Fig:OrthogonallAmplitudes1}
\end{figure}

\begin{figure}[htb]
     \centering
         \includegraphics[width=0.45\textwidth]{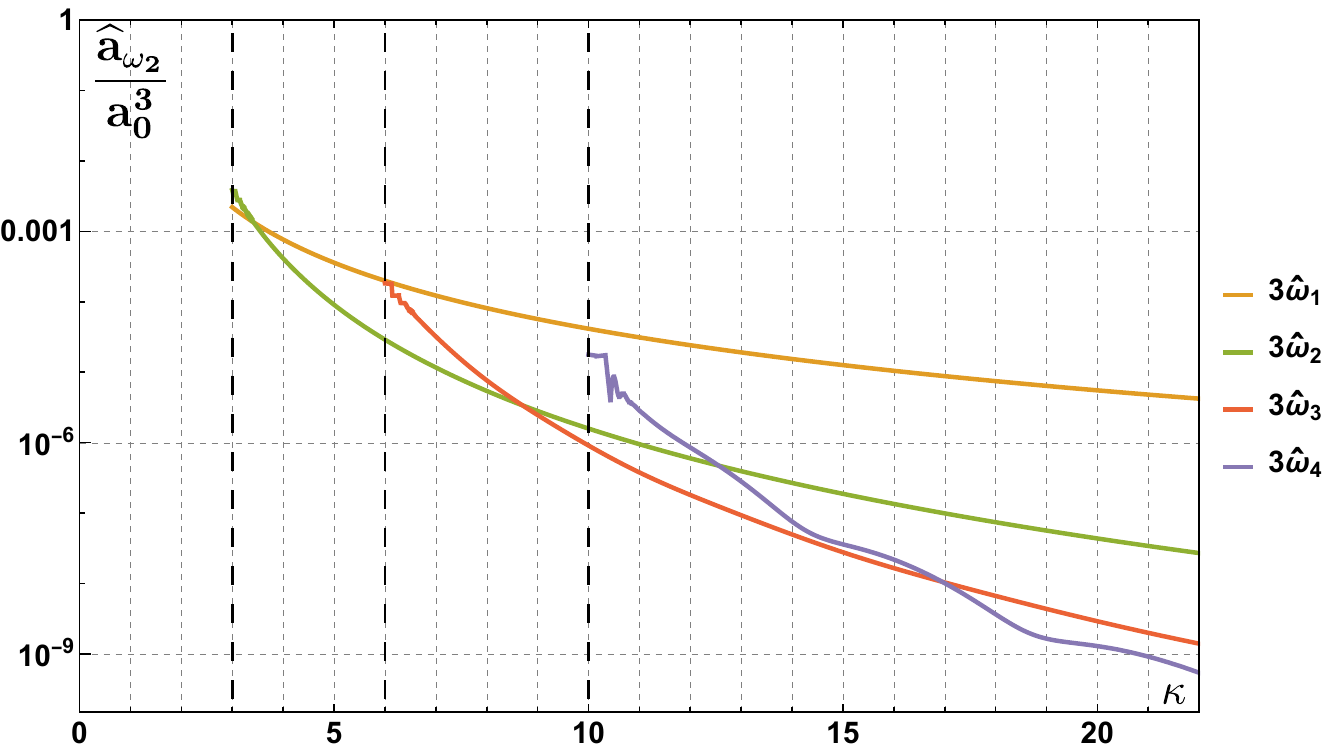}
         \vspace{-0.3cm}
         \caption{Radiation amplitudes  associated with the frequencies $ 3\widehat{\omega}_1,\dots 3\widehat{\omega}_4$ in the orthogonal  channel. The dashed lines indicate the values of $\kappa$ for which a new shape mode arises in the spectrum of $\mathcal{H}_{22}$.}
         \label{Fig:OrthogonallAmplitudes2}
\end{figure}

Next we focus on obtaining the radiation amplitudes associated with each frequency as a function of the coupling constant $\kappa$, depending on which orthogonal mode is initially activated. For this we must use \eqref{FinalRadiationFirstField}--\eqref{FinalRadiationSecondField}, which must necessarily be evaluated numerically, showing the results in Figures~\ref{Fig:LongitudinalAmplitudes}--\ref{Fig:OrthogonallAmplitudes2}.
More specifically, the real part of the result obtained in \eqref{FinalRadiationFirstField} must be taken for the longitudinal channel amplitudes and the imaginary part of the result found in \eqref{FinalRadiationSecondField} for the orthogonal channel amplitudes.

In Figure~\ref{Fig:LongitudinalAmplitudes} we show the behavior of longitudinal radiation amplitudes when the first five orthogonal modes are activated separately.
As we can see, as $\kappa$ grows, the radiation amplitudes get smaller, which can be explained by the fact that higher frequencies are more difficult to trigger, and as $\kappa$ grows, the gap between $2\widehat {\omega}_i$ and the threshold $\overline{\omega}^c_0$ also increases.
Furthermore, it can be seen that the radiation emitted when we excite $\widehat{\eta}_{D,0}$ is much bigger than when we trigger  higher shape modes. 
In addition to all this, it is important to point out that, for large values of $\kappa$, the amplitudes associated with the higher modes are smaller than those corresponding to the first modes. 
For example, for $\kappa>20$, $\overline{a}_{2\widehat{\omega}_0}>\overline{a}_{2\widehat{\omega}_1}>\overline{a}_{2\widehat{\omega}_2}>\overline{a}_{2\widehat{\omega}_3}>\overline{a}_{2\widehat{\omega}_4}$. 
Another notable phenomenon is that the radiation associated with the frequency $2\widehat{\omega}_0$ begins when $\kappa=6$. 
As mentioned above, this is because when $\kappa<6$ the aforementioned frequency is not embedded in the continuous part of the spectrum of the longitudinal channel (see Figure \ref{Fig:LongitudinalEspectrum}).

On the other hand, in Figure \ref{Fig:OrthogonallAmplitudes1} we can see the graphs corresponding to the amplitudes of the orthogonal radiation emitted in $\overline{\omega}+\widehat{\omega}_j$, $j=1,\dots,4$, as a function of $\kappa$.
In this case, these frequencies only emit radiation in a certain range of values of $\kappa$.
In other words, the coupling between $\widehat{\eta}_{D,j}$ and $\overline{\eta}_D$ only produces radiation at $\overline{\omega}+\widehat{\omega}_j $ for a value of $\kappa$ that is greater than the minimum value for which the orthogonal shape mode arises and less than the critical value for which this frequency no longer belongs to the continuous part of the spectrum of the second channel.
Note also that the amplitude of the radiation diverges near the value of $\kappa$ where there is a resonance between $\overline{\omega}+\widehat{\omega}_i$ and the threshold value of $\widehat{\omega}_c$. 
These resonance structures must be addressed by other analytical methods due to the fact that these limits are outside the range of validity where our perturbative approach works well.
Furthermore, the first orthogonal shape mode cannot trigger radiation in the second field component because this frequency is always below the continuous spectrum.

Finally,  Figure \ref{Fig:OrthogonallAmplitudes2} contains the graphs of the amplitudes corresponding to the radiation emitted at frequencies $3\widehat{\omega}_1,\dots 3\widehat{\omega}_4$. 
Note that, as in Figure \ref{Fig:LongitudinalAmplitudes}, for large values of $\kappa$ the amplitudes associated with the higher modes are smaller that those corresponding to the first shape modes.  In fact, in this case the radiation amplitudes are much smaller than in the previous cases. 
This phenomenon can be explained taking into account that the higher frequencies are more difficult to excite, as can be seen in Figures \ref{Fig:LongitudinalEspectrum} and \ref{Fig:OrthogonalSpectrum}: the frequencies $3\widehat{\omega }_j $ are much larger than $2\widehat{\omega}_j$ and $\widehat{\omega}_j+\overline{\omega}$.
Note also that the orthogonal channel radiation terms  are proportional to $a_0^3$, making them much smaller than the radiation propagated in the longitudinal channel, which is proportional to $a_0^2$.

In the next section we will compare all the analytical results that we have just developed with those obtained through numerical simulations.


\section{Numerical analysis}\label{Section4} 

Once we have developed the perturbative method for the problem we are analyzing in the preceding section, it now seems reasonable to compare the results obtained there with those that arise when the field equations \eqref{FieldEqn1}--\eqref{FieldEqn2} are solved numerically. To carry out these simulations, the aforementioned nonlinear partial differential equations have been discretized using an explicit fourth-order finite difference algorithm implemented with fourth-order Mur boundary conditions \cite{AlonsoIzquierdo2021} in the spatial interval $x\in (- 100,100)$ for a time $0<t<1200$. The initial configuration is determined by \eqref{InitialHypothesis}, that is, the same one used in the perturbation approach developed in Section~\ref{Section3}.
Specifically, the simulations have been run for initial configurations for which one of the first three orthogonal eigenmodes has been excited and for various initial amplitudes. To study the radiation emitted by the wobbling kink and its internal vibration, the Fast Fourier Transform algorithm has been implemented at several points on the real axis to obtain the spectral data. 
This analysis has been carried out at points far from both the center of the kink ($x_B$) and the points where the shape modes have their maxima. 
For $\overline{\eta}$ the maximum is 
\begin{equation}
x_M=\ln(1+\sqrt{2}),
\end{equation}
and  for the first three orthogonal modes $\widehat{\eta}_{D,0}$, $\widehat{\eta}_{D,1}$ and $\widehat{\eta}_{D,2}$ the maxima are, respectively,
\begin{eqnarray}
x_{M0}\!\!&\!\! = \!\!&\!\!x_0=0,
\\ 
x_{M1 }\!\!&\!\! = \!\!&\!\!\arctanh\left(\sqrt{\frac{2}{\sqrt{8\kappa+1}-1}}\, \right), 
\\
x_{M2}\!\!&\!\! = \!\!&\!\!\frac{1} {2}\arccosh \left(\frac{-5+4\kappa+\sqrt{1+8\kappa}}{4(2+\kappa-\sqrt{1+8\kappa})}\,\right),
\end{eqnarray}

For ease of presentation, this section will be organized as follows: in Section~\ref{Section4.1} we will check whether the assumption of a constant amplitude associated with the excited orthogonal shape mode is true in the perturbative regime.  
In Section~\ref{Section4.2} we will discuss how the initially triggered shape mode couples with longitudinal vibration mode. 
In Section~\ref{Section4.3} the radiation emitted by the kink along the longitudinal and orthogonal channels will be analyzed. 
In Section~\ref{Section4.4} a study of the coupling between orthogonal modes will be addressed. Finally, in Section~\ref{Section4.5} we will present an analytical explanation of the energy loss of the first orthogonal shape mode when the initial amplitude is increased.

\subsection{Hypothesis validation (first order in $a_0$)}\label{Section4.1}

\begin{figure}[htb]
     \centering
         \includegraphics[width=0.4\textwidth]{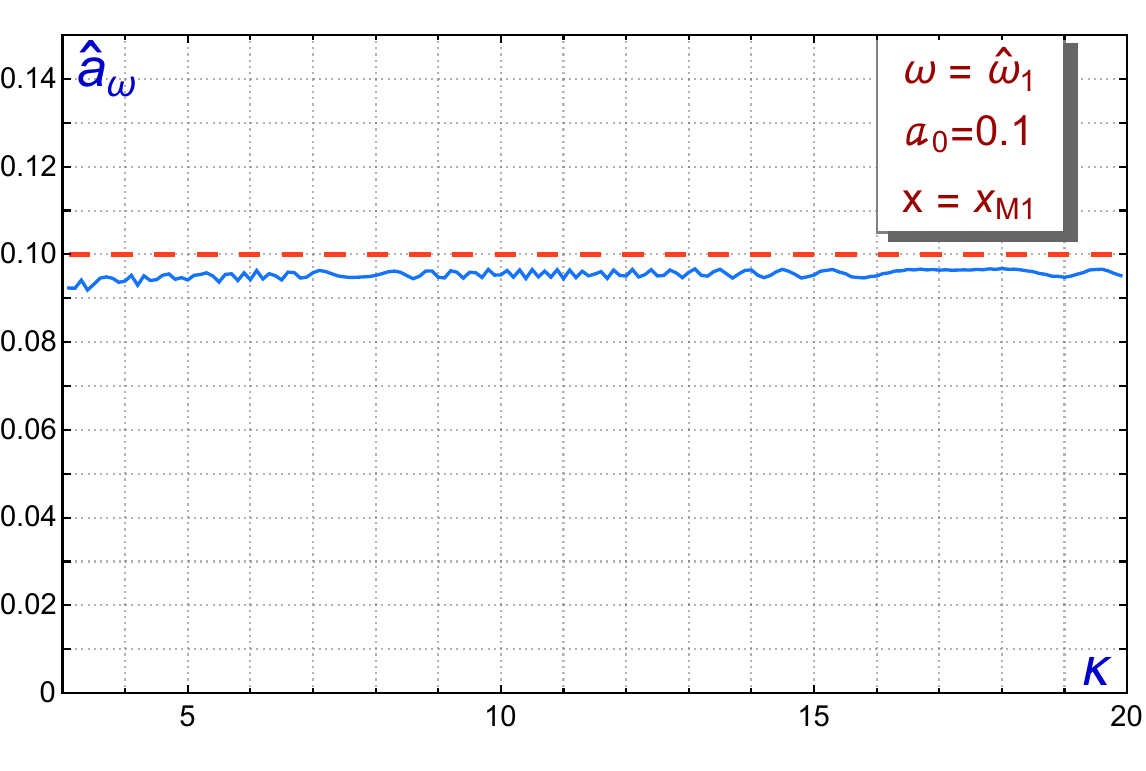}\qquad
         \includegraphics[width=0.4\textwidth]{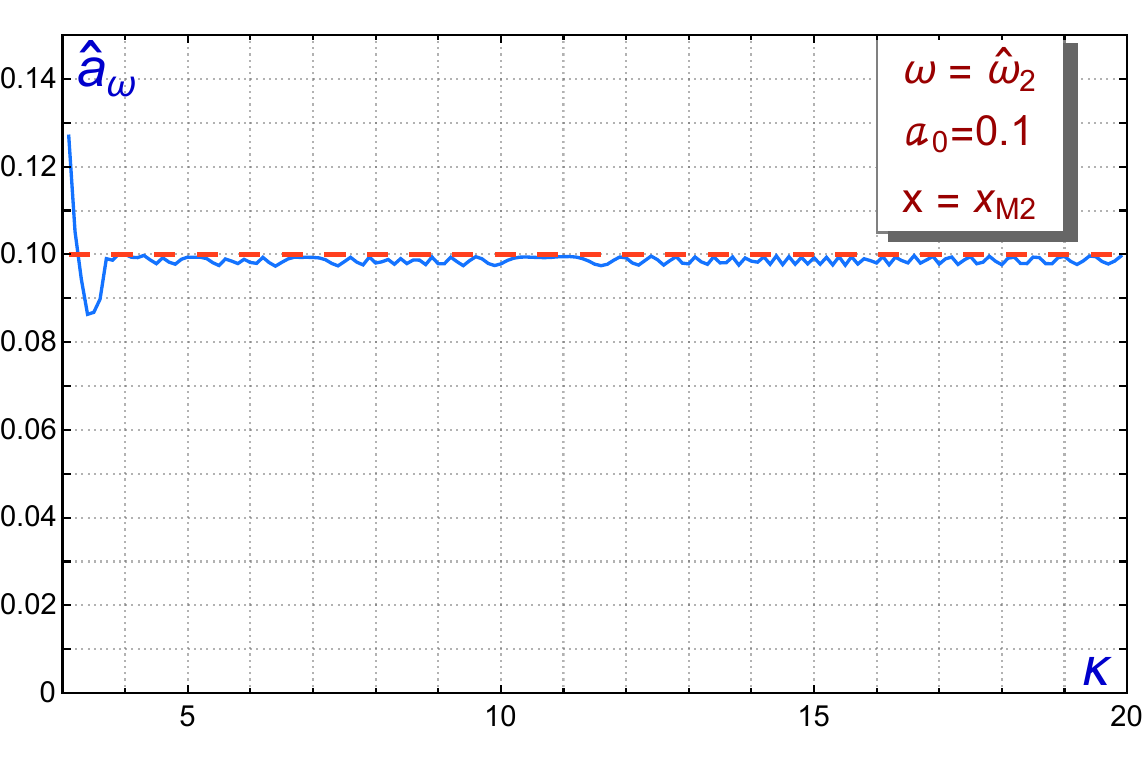}
     \caption{Vibration amplitudes (blue curve) associated with  $\widehat{\eta}_{D,1}$ (first drawing) and $\widehat{\eta}_{D,2}$ (second drawing) as a function of $\kappa$  when the second and third shape modes are initially activated. The red dashed line corresponds to the analytical hypothesis \eqref{OthogonalAmplitudeEvolution}.}
     \label{Fig:Mode0-1-2-OrthogonalAmplitude}
\end{figure}

In Section~\ref{Section2} it was assumed that, when an orthogonal mode is activated, its associated amplitude remains essentially constant \eqref{OthogonalAmplitudeEvolution}.
In Figure \ref{Fig:Mode0-1-2-OrthogonalAmplitude} it can be seen that this hypothesis agrees quite well with the numerical results for various typical values, specifically  for $\widehat{\eta}_{D,1}$ (first drawing) and $\widehat {\eta}_{D,2}$ (second drawing) with $a_0\approx \mathcal{O}(0.1)$.

\begin{figure}[htb]
     \centering
     \includegraphics[width=0.4\textwidth]{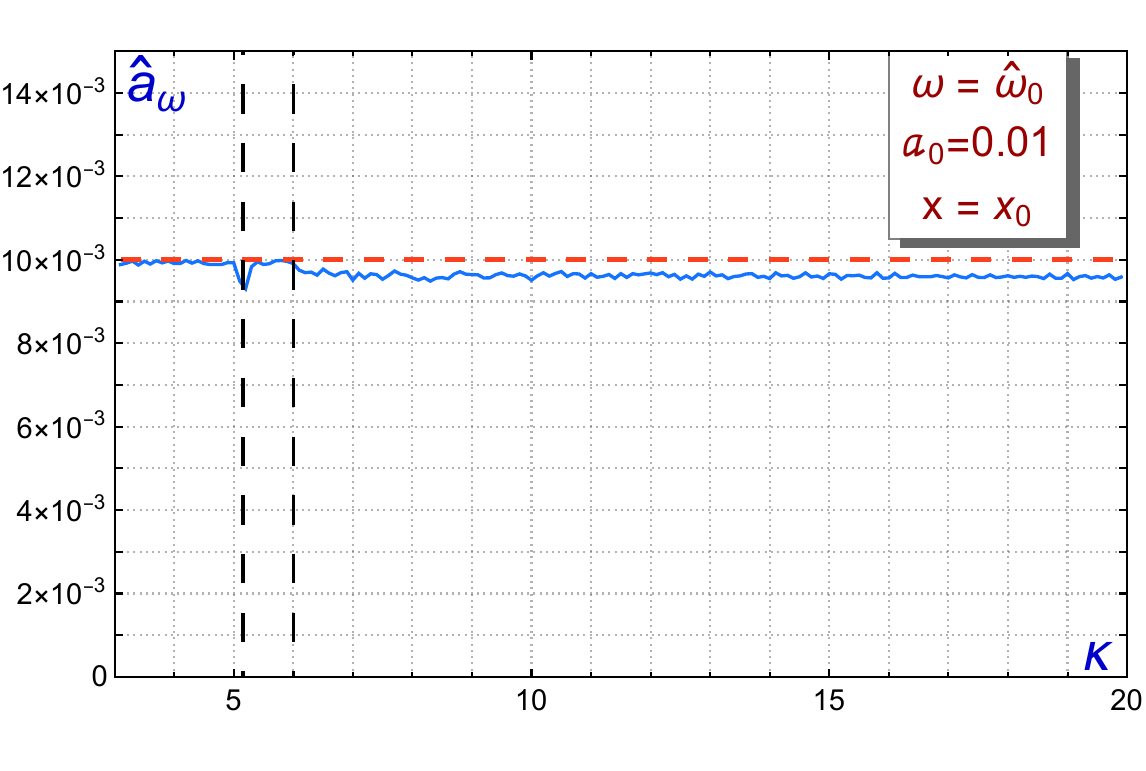}\qquad
         \includegraphics[width=0.4\textwidth]{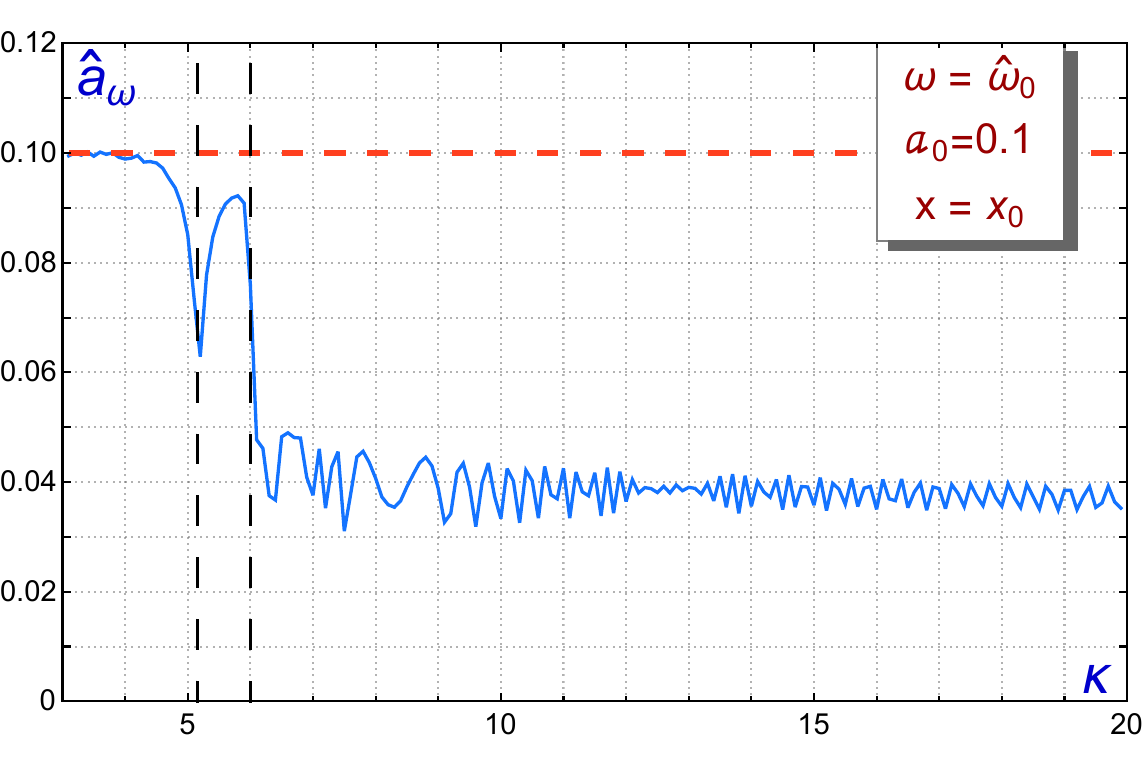}
 \qquad
          \includegraphics[width=0.4\textwidth]{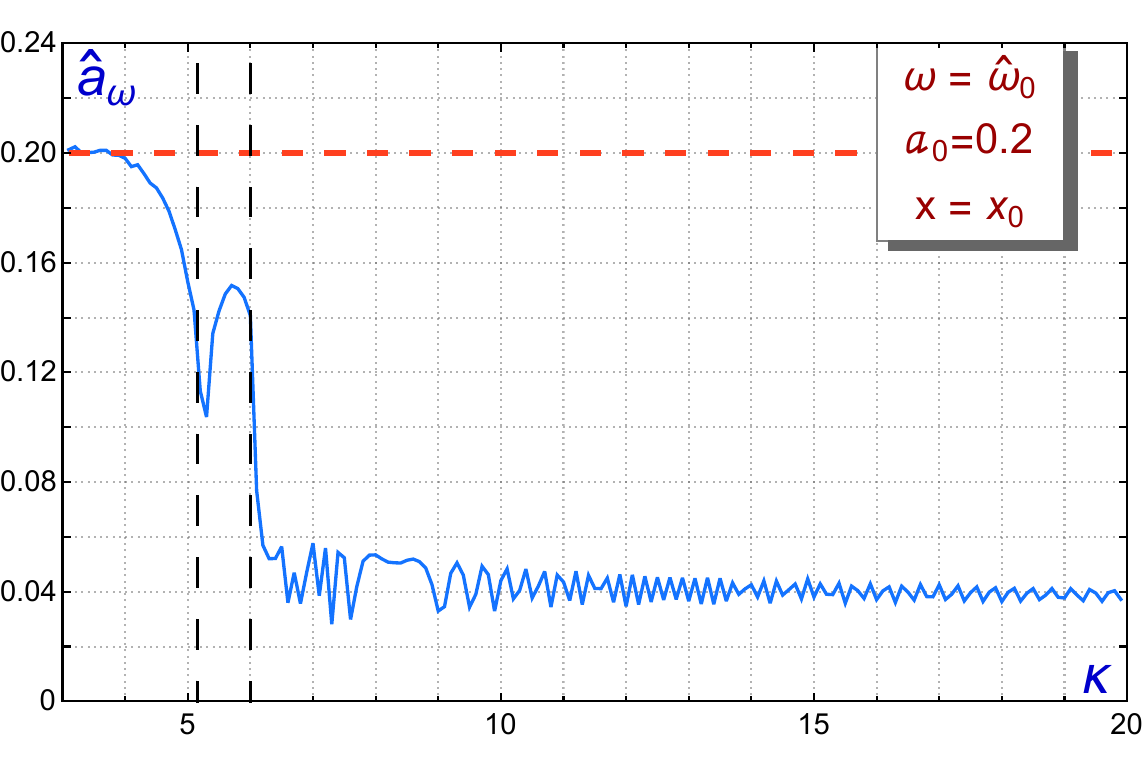}
     \caption{Vibration amplitudes associated with $\widehat{\eta}_{D,0}$ as a function of $\kappa$,  for $a_0=0.01, 0.1$ and $0.2$.}
     \label{Fig:Mode0rthogonalAmplitude}
\end{figure}

However, in Figure~\ref{Fig:Mode0rthogonalAmplitude}  it can be seen that this hypothesis works fine for amplitudes of order $a_0\approx\mathcal{O }(0.01)$ for $\widehat{\eta}_{D,0}$, but fails when considering the simulations  with $a_0\approx\mathcal{O }(0.1)$.
Indeed, a large decrease in the initial amplitude can be observed for $\kappa>6$, which is the regime in which the kink is capable of emitting radiation with frequency $2\widehat{\omega}_0$. This phenomenon can be explained if we take into account that for large values of $t$ part of the vibration energy is dissipated in the form of radiation. 
In fact, the amplitude of the radiation emitted when $\widehat{\eta}_{D,0}$ is excited is much larger than when higher shape modes are activated (see Figure~\ref{Fig:LongitudinalEspectrum}). This is also the reason why this decay is much smaller when we consider the simulations performed for $\widehat{\eta}_{D,1}$ and $\widehat{\eta}_{D,2}$ (see Figure~\ref{Fig:Mode0-1-2-OrthogonalAmplitude}).
Since the radiation emitted when we consider $\widehat{\eta}_{D,1}$ is greater than that emitted when we consider  
$\widehat{\eta}_{D,2}$, then the observed decrease in $ a_0$ in this last case will be less than for the second shape mode (see  Figure~\ref{Fig:Mode0-1-2-OrthogonalAmplitude}). A decay law for this amplitude will be discussed in Section~\ref{Section4.5} taking into account the radiation emitted by the wobbling kink.
In addition to all that has been mentioned above, analyzing the Figures~\ref{Fig:LongitudinalEspectrum} and \ref{Fig:Mode0rthogonalAmplitude}, an additional decrease in the values of $a_0$ can be observed  for $\kappa\approx 5.15$, which is the value for which $2\widehat{\omega}_0=\overline{\omega}$.

\subsection{Amplitude of the longitudinal shape mode (second order in $a_0$)}\label{Section4.2}

\begin{figure}[htb]
     \centering
         \includegraphics[width=0.4\textwidth]{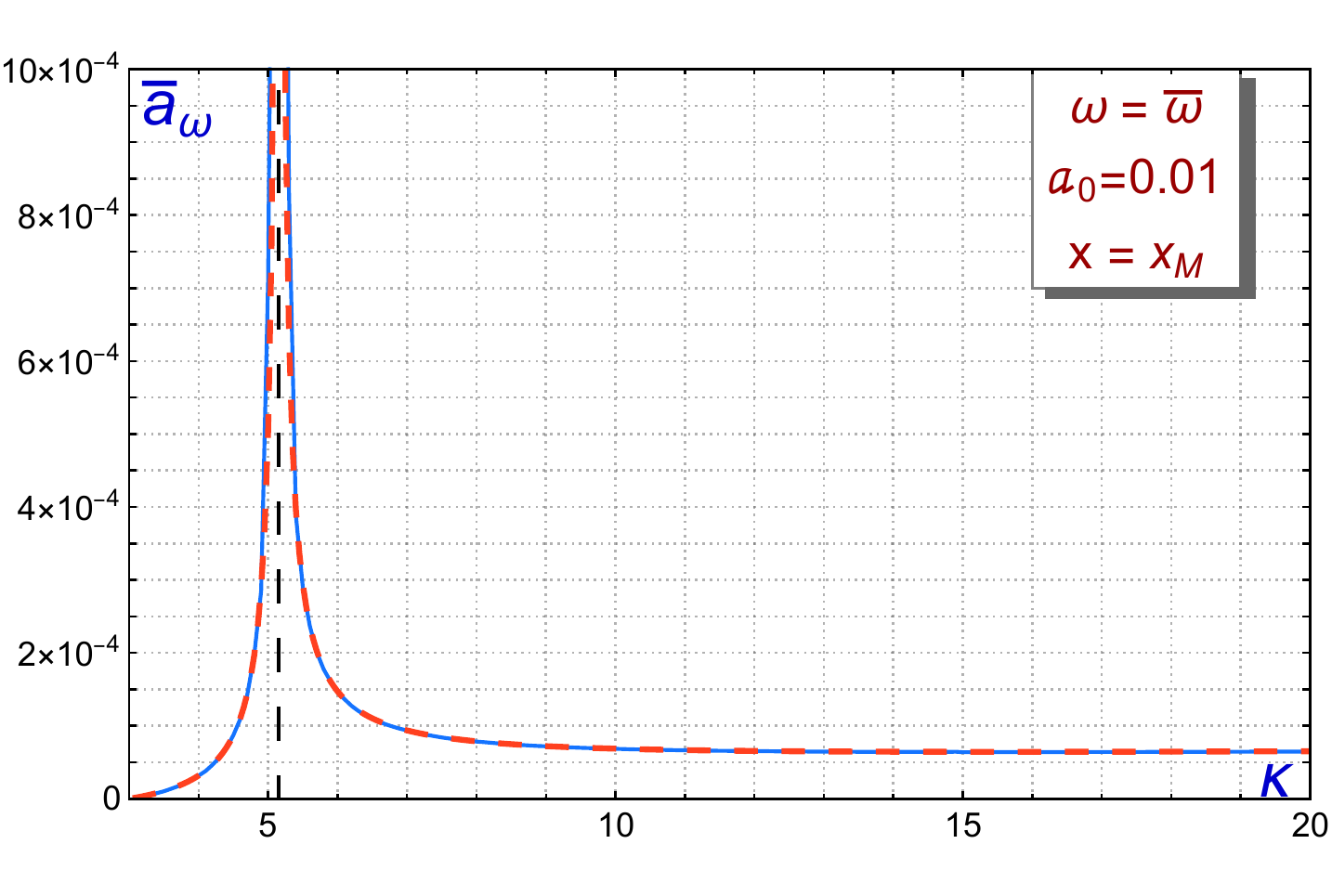}
\qquad
         \includegraphics[width=0.4\textwidth]{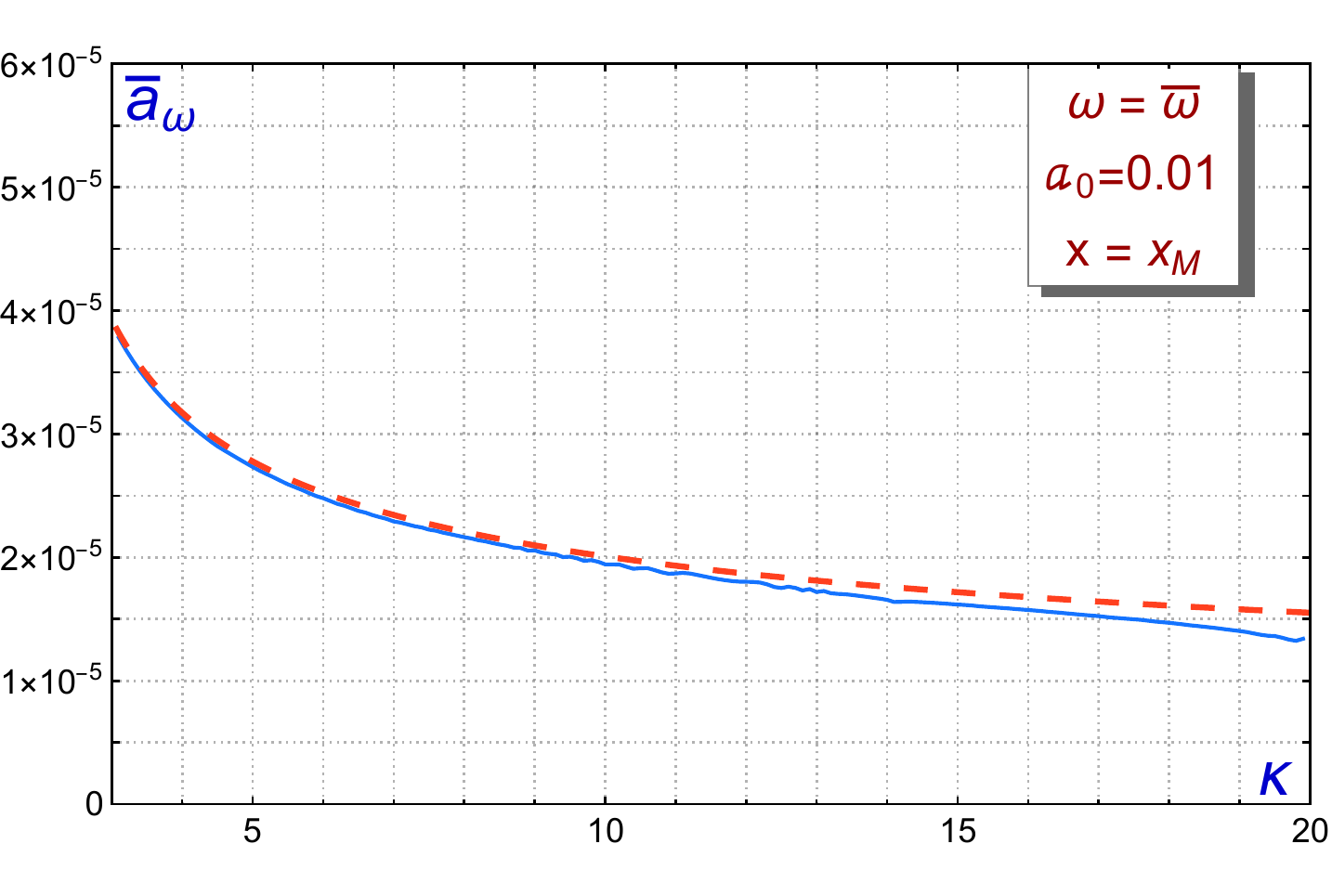}
\qquad
         \includegraphics[width=0.4\textwidth]{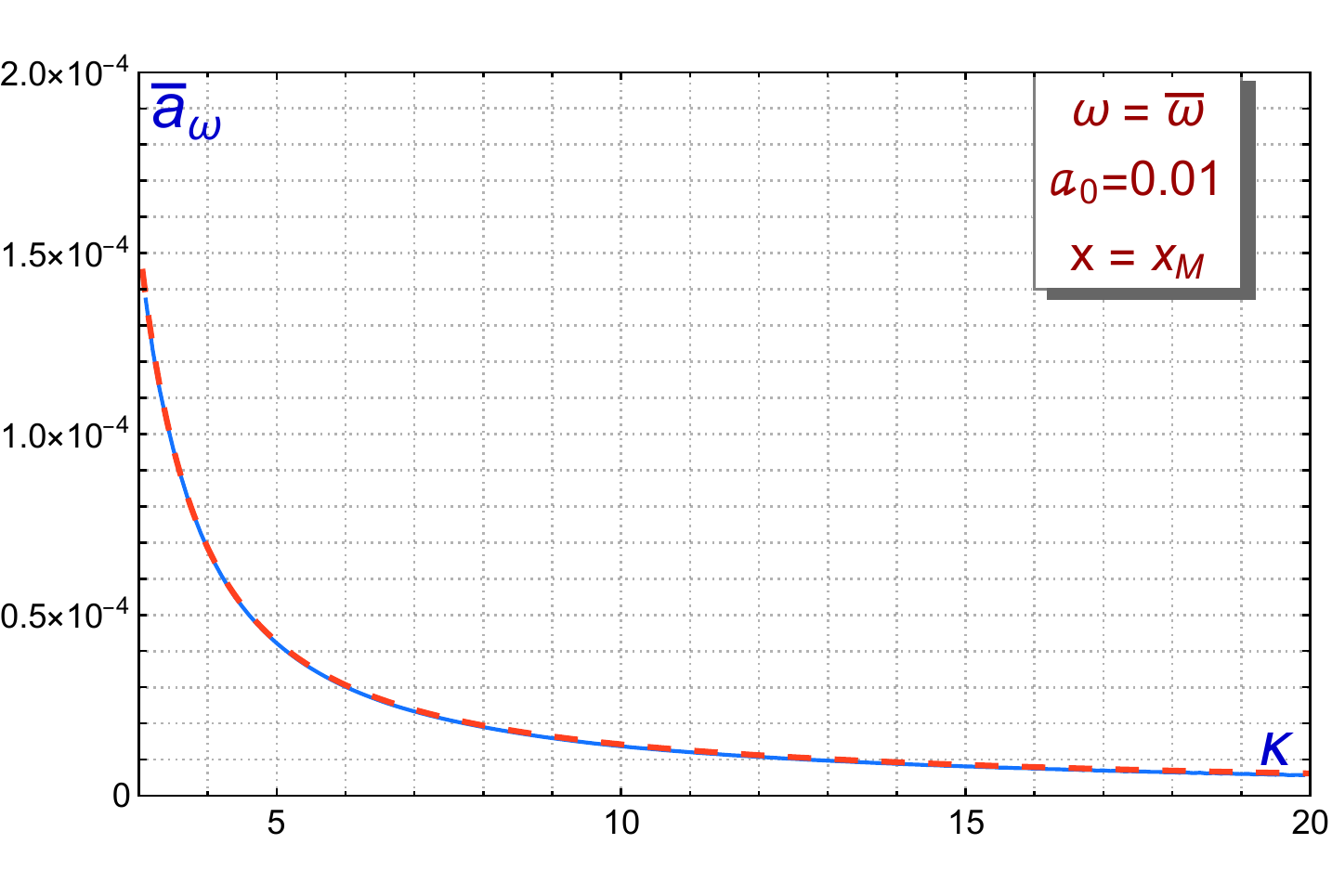}
     \caption{Vibration amplitudes (blue curves) of the longitudinal mode with frequency $\overline{\omega}$ in \eqref{LongitudinalAmplitudeEvolution} when $\widehat{\eta}_{D,0}$ (first drawing), $\widehat {\eta}_{D,1}$ (second drawing) and $\widehat{\eta}_{D,2}$ (third drawing) are triggered, always with $a_0=0.01$. The red dashed lines correspond to the analytical estimate.}
     \label{Fig:Mode0-1-2-LongitudinalAmplitude}
\end{figure}

In Figure~\ref{Fig:Mode0-1-2-LongitudinalAmplitude} it can be seen that the analytical estimate made in Section~\ref{Section2} is fully consistent with the numerical simulations we have developed.    
It is worth mentioning that in the first drawing of Figure~\ref{Fig:Mode0-1-2-LongitudinalAmplitude} it is possible to observe a resonance for the value $\kappa=\frac{165}{32}\approx5.15$, which coincides with the decrease in the amplitude of the shape mode observed in the previous section, specifically in the first drawing of Figure~\ref{Fig:Mode0-1-2-OrthogonalAmplitude} and in Figure~\ref{Fig:Mode0rthogonalAmplitude}.
This means that a large amount of energy  is transferred from the shape mode to the longitudinal mode for this particular value of the coupling constant, where it happens that $2\widehat{\omega}_0=\overline{\omega}$. 
In fact, we will see in Section~\ref{Section4.4} that part of this energy is also transferred to $\widehat{\eta}_{D,2}$ when $\kappa=6$.

It can also be seen that, for large values of $\kappa$, when considering higher shape modes, the amplitude of the longitudinal shape mode becomes smaller and smaller.
Another remarkable phenomenon is that, for $\kappa>>1$, the amplitude of the shape mode in \eqref{LongitudinalAmplitudeEvolution} can be approximated as 
\begin{equation}
   \overline{a}_{\overline{\omega}}\approx \frac{a_0^2\, \widehat{B}_{jj}}{2\, \overline{C}\ \overline{\omega}^2}
=\frac{a_0^2\, \widehat{B}_{jj}}{4}.
\end{equation}

\subsection{Radiation amplitudes (second-third order in $a_0$)}\label{Section4.3}

\begin{figure}[htb]
     \centering
         \includegraphics[width=0.40\textwidth]{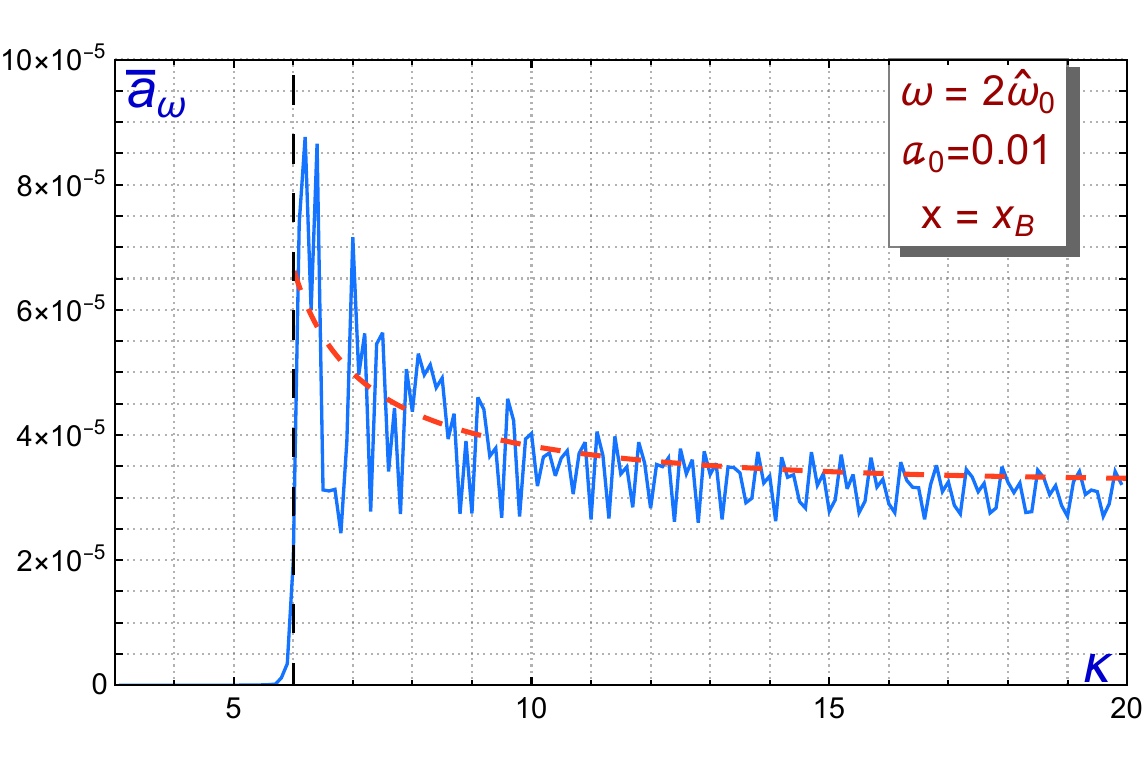} 
         \includegraphics[width=0.40\textwidth]{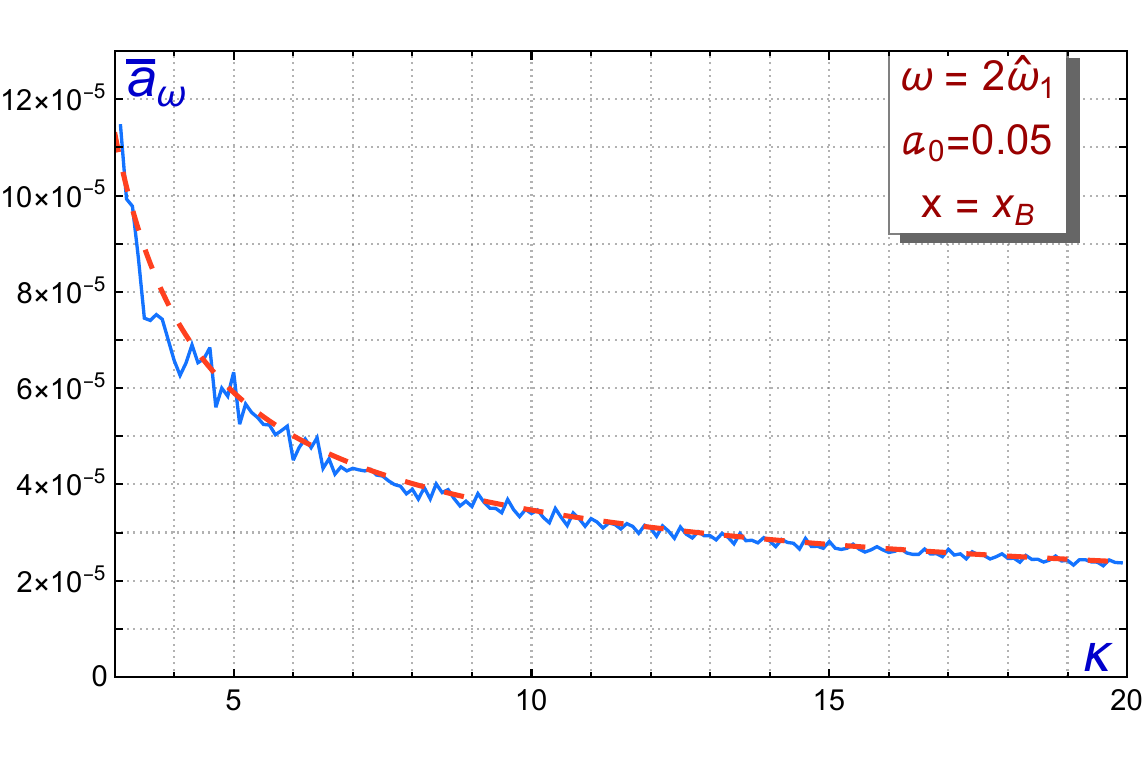} 
         \includegraphics[width=0.40\textwidth]{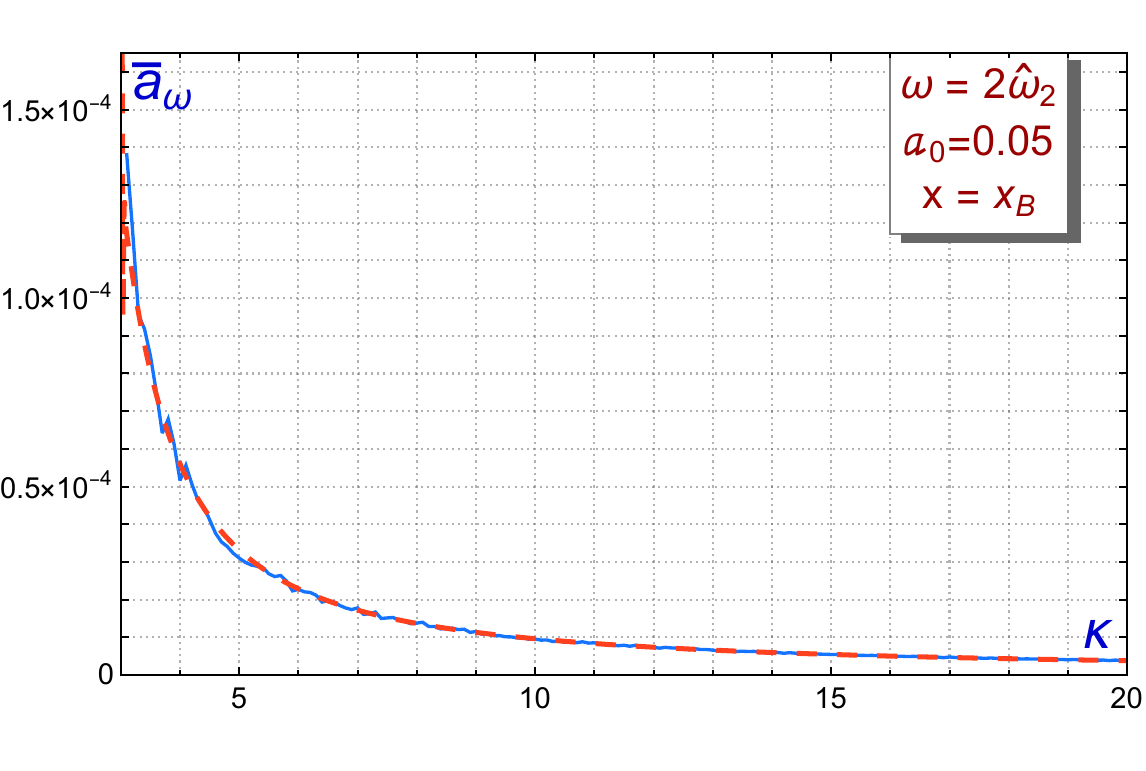}
     \caption{Radiation amplitudes (blue curves) emitted   at the frequency $2\widehat{\omega}_j$ by the kink when we trigger $\widehat{\eta}_{D,0}$ (first drawing), $\widehat{\eta}_{D,1}$ (second drawing) and $\widehat{\eta}_{D,2}$ (third  drawing). 
     The red dotted lines represent the analytical prediction \eqref{FinalRadiationFirstField}.}
     \label{Fig:RadiationLongitudinalNumeric}
\end{figure}

In Section~\ref{Section3} we show  that the kink is capable of emitting radiation  when we trigger at least one of its shape modes. 
In this section we will compare  the  radiation amplitudes obtained by numerical simulations with the theoretical predictions for the longitudinal  and  orthogonal channels  separately. 
Firstly we will focus on studying the behavior of longitudinal radiation.
Note that in the first drawing of  Figure~\ref{Fig:RadiationLongitudinalNumeric} it can be seen that the excited kink cannot emit radiation with a frequency $2\widehat{\omega}_0$ when $\kappa<6$. 
This phenomenon is due to the  fact that the frequency $2\widehat{\omega}_0$ is only embedded into the continuous spectrum of the longitudinal mode when $\kappa>6$ (see Figure~\ref{Fig:LongitudinalEspectrum}). 
On the other hand, since $2\widehat{\omega}_1$ and $2\widehat{\omega}_2$ are always included in the continuous spectrum, radiation associated with these frequencies will always be found, as can be seen in the second and third drawings of Figure~\ref{Fig:RadiationLongitudinalNumeric}, in which it can also be seen that the numerical results coincide very well with the analytical ones.
Note also that the highest radiation amplitude is the one associated with $\widehat{\eta}_{D,0}$ and that, when  considering higher shape modes, the radiation amplitudes corresponding to the frequency $2\widehat{\omega}_j$ become smaller, as predicted in Section~\ref{Section3}.

\begin{figure}[htb]
     \centering
         \includegraphics[width=0.4\textwidth]{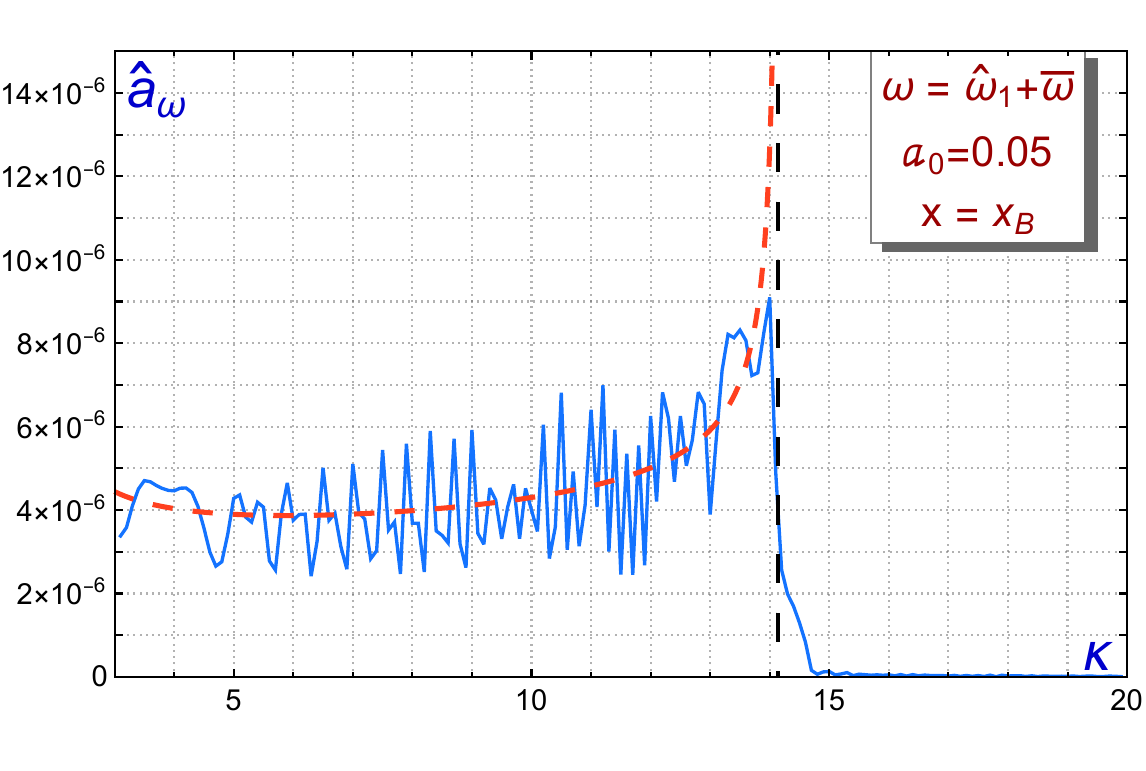}
         \qquad
         \includegraphics[width=0.4\textwidth]{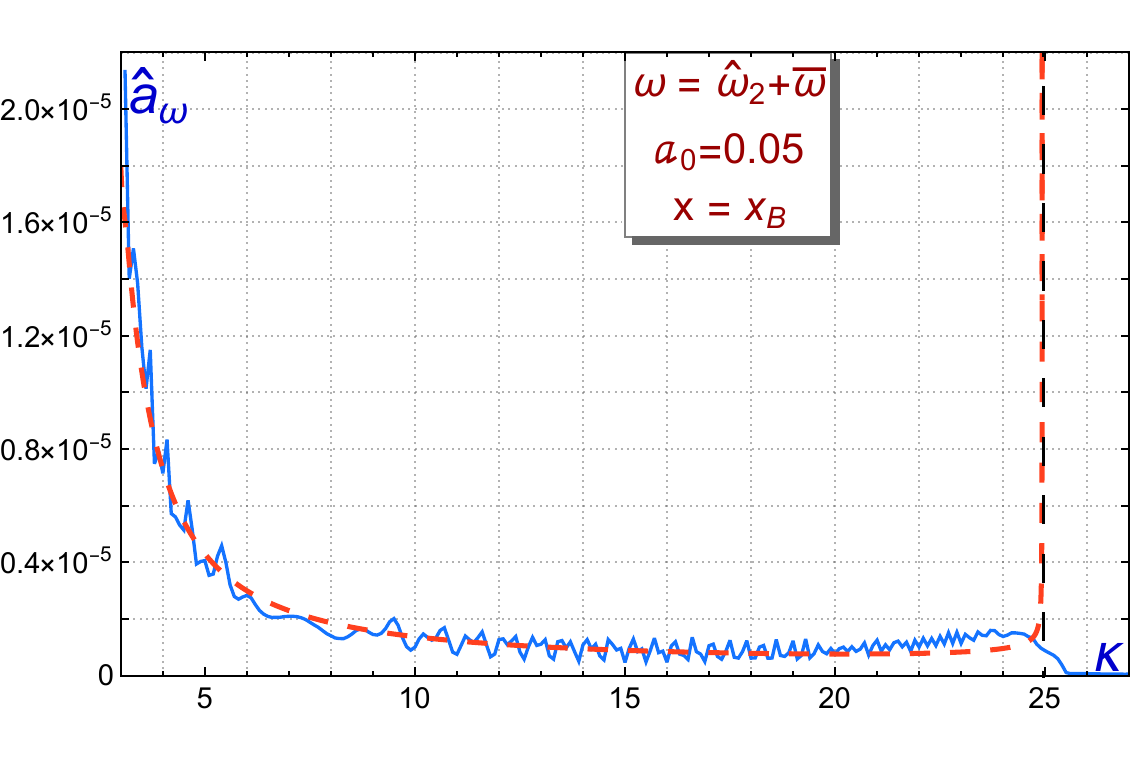}
    \caption{Radiation amplitudes (blue curve) emitted  at frequency $\widehat{\omega}_j+\overline{\omega}$ by the kink  when we trigger the orthogonal shape modes $\widehat{\eta}_{D,1}$ (first drawing) and  $\widehat{\eta}_{D,2}$ (second drawing). 
    The red dotted line represents the analytical prediction of formula \eqref{FinalRadiationSecondField}.}
     \label{Fig:RadiationOrthogonalNumeric}
\end{figure}

The behavior of orthogonal radiation is shown in Figure \ref{Fig:RadiationOrthogonalNumeric}.
From the simulations performed for $\widehat{\eta}_{D,1}$, shown in the first drawing of Figure~\ref{Fig:RadiationOrthogonalNumeric}, it can be seen that no radiation is found in the orthogonal channel for $\kappa>14.14$ at frequency $\widehat{\omega}_1+\overline{\omega}$.
In this range, the aforementioned  frequency  is less than the threshold  value of $\widehat{\omega}^c_0$. 
It can also be observed the existence of a resonance close to this particular value of $\kappa$, which agrees with the result obtained with the equation \eqref{FinalRadiationSecondField}.
Furthermore, on the second drawing of Figure~\ref{Fig:RadiationOrthogonalNumeric} it can be observed that a similar phenomenon occurs for the simulation performed with $\widehat{\eta}_{D,2}$: the kink stops emitting radiation in the range $\kappa >24.93$, which agrees with the analytical prediction made in Section~\ref{Section3}.
Finally, it is worth mentioning that no radiation was found for $\widehat{\omega}_0+\overline{\omega}$ because this frequency is less than  $\widehat{\omega}_0^c$.

The graphs corresponding to the radiation emitted at $\omega=3\widehat{\omega}_j$ will not be shown, since from Figure~\ref{Fig:OrthogonallAmplitudes2} it can be seen that these magnitudes are too small, which makes it difficult to  observe them adequately from the data extracted from numerical simulations. 
This fact can be explained because higher frequencies are more difficult to excite and higher nonlinearities end up exciting frequencies close to the continuous frequency threshold.

\subsection{Amplitude of other orthogonal shape modes (third order in $a_0$)}\label{Section4.4}

In Section~\ref{Section3} it was found that exciting a certain orthogonal mode also activates all other orthogonal modes that have the same parity.
This means that if, for example, $\widehat{\eta}_{D,0}$ is excited, then $\widehat{\eta}_{D,2}, \widehat{\eta}_{ D, 4}, \widehat{\eta}_{D,6},\dots$ will have a non-zero amplitude.
In fact, the analytical expression that describes this event is given by the formula \eqref{OrthogonalAmplitudes3}.
In this section we will analyze some examples of this phenomenon.

\begin{figure}[htb]
     \centering
         \includegraphics[width=0.395\textwidth]{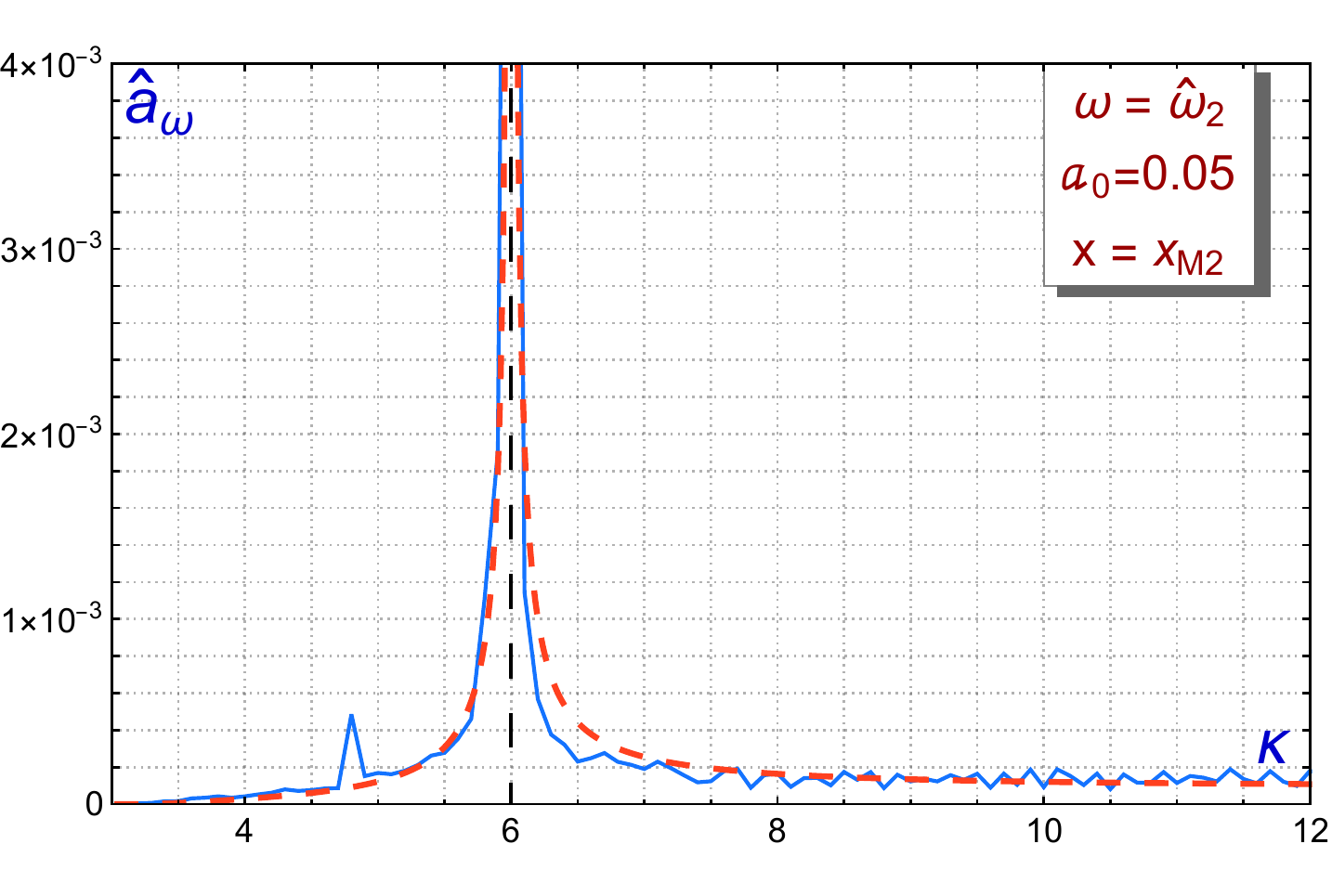}  
         \includegraphics[width=0.395\textwidth]{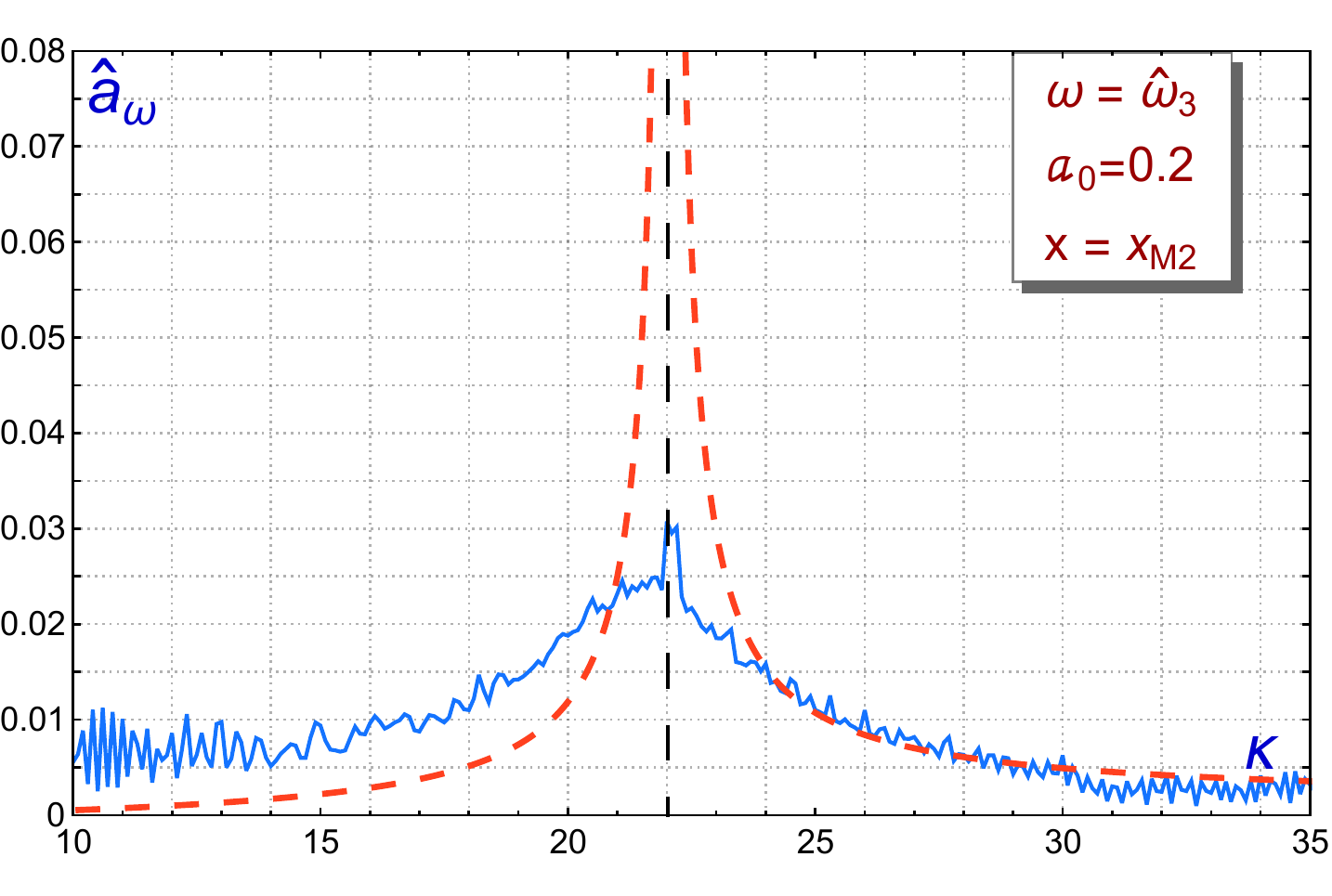} 
         \includegraphics[width=0.395\textwidth]{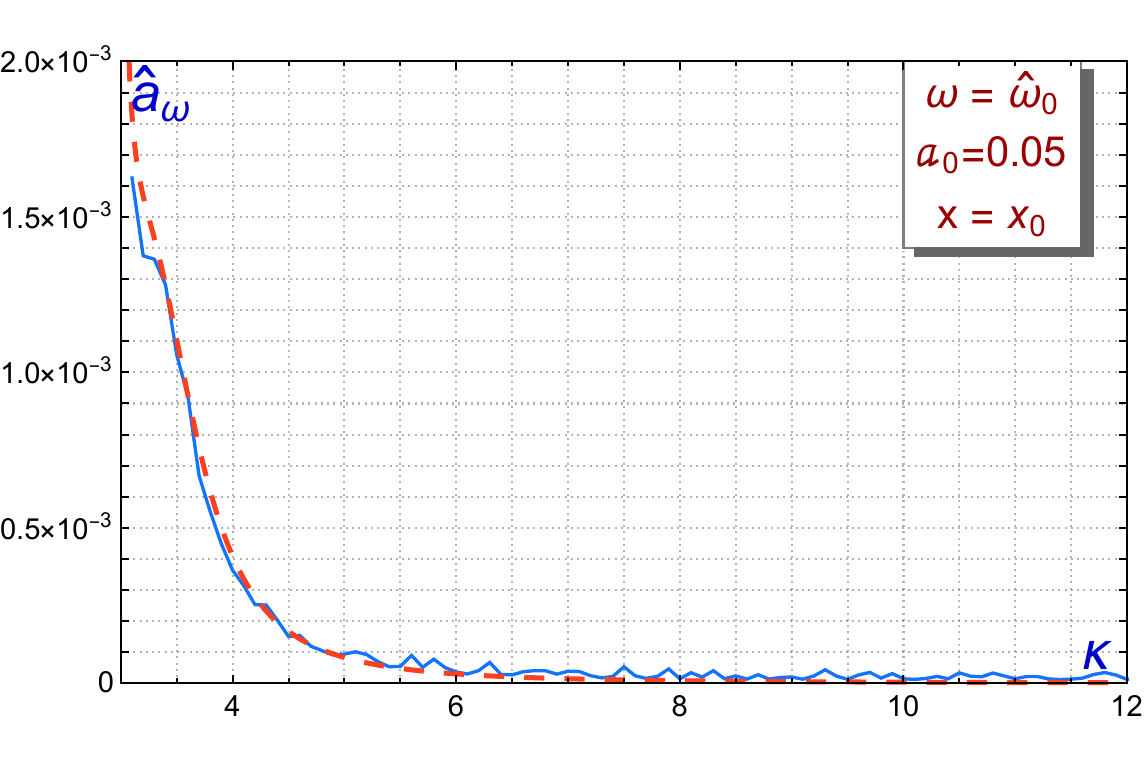}
     \caption{Numerical vibration amplitudes (blue curves) associated with excitation of the shape modes $\widehat{\eta}_{D,2}$ (first drawing), $\widehat{\eta}_{D,3}$ (second drawing) and $\widehat{\eta}_{D,0}$ (third drawing) when $\widehat{\eta}_{D,0}$, $\widehat{\eta}_{D,1}$ and $\widehat{\eta}_{D,2}$ were initially excited, respectively. 
     The red dashed lines correspond to the analytical prediction \eqref{OrthogonalAmplitudes3}.}
     \label{Fig:CouplingModes}
\end{figure}

In Figure~\ref{Fig:CouplingModes} the theoretical (red curves) and numerical (blue curves) results obtained for the amplitudes associated with $\widehat{\eta}_{D,2}$, $\widehat{\eta}_{D,3 } $ and $ \widehat {\eta}_{D,0}$, as a function of the coupling constant $\kappa$, are shown together. It is obvious that, depending on the case, the similarity between both types of results is better or worse.
     It is important to point out that several divergences appear in the graphs, which are due to different resonances between the frequencies involved in the calculation carried out to obtain the behavior of these amplitudes.
     For example, in the first drawing in Figure \ref{Fig:CouplingModes} there is a divergence at $\kappa=6$ that can be explained by the fact that, for this specific value of the coupling constant, there is a resonance between $ \widehat{\omega}_2$ and $3\widehat{\omega}_0$, as can be seen on the left side of Figure~\ref{Fig:OrthogonalSpectrum}.

On the other hand, in the second of the graphs in Figure~\ref{Fig:CouplingModes} it is possible to identify the amplitude associated with $\widehat{\eta}_{D,3}$ when $\widehat{\eta}_{D,1}$ is triggered.  This amplitude has been plotted for $\kappa>10$, which is the range where this orthogonal shape mode arises. 
In this case we can notice the presence of a divergence at $\kappa\approx 22.02$, which is due to a resonance between $\widehat{\omega}_3$ and $\widehat{\omega}_1+\overline{\omega}$.
 Finally, in the third drawing of Figure~\ref{Fig:CouplingModes} it can be seen that we cannot appreciate any type of coupling between frequencies, but it can be verified that there is a divergence at $\kappa=3$, which is the value for which $\widehat{\omega}_0=0$.

\subsection{Decay law for the orthogonal  mode amplitude }\label{Section4.5}

 In Section~\ref{Section4.1} it was shown that if we set $\widehat{\eta}_{D,0}$ with $a_0\approx\mathcal{O}(0.01)$, then $a_0$ remains constant, as assumed in \eqref{OthogonalAmplitudeEvolution}.
 However, this is not true when considering higher values of the initial amplitude and the  $\kappa>6$ regime.
In this section, this energy loss will be quantified taking into account the radiation emitted by the wobbling kink  for this specific case.
 
Next we will show that a decay law can be calculated analytically using reasoning similar to that used in \cite{Manton1997} and \cite{AlonsoIzquierdo2023}.
In Section~\ref{Section3} we assumed that if we initially trigger $\widehat{\eta}_{D,0}$, its corresponding amplitude can be approximated as
\begin{equation}
    (\widehat{a}_{0})_{tt}+ \widehat{\omega}_0^2\  \widehat{a}_0\approx 0,
\end{equation}
which implies that the first orthogonal shape mode behaves as a harmonic oscillator at each point on the real axis. 
Thus, the  energy density  can be written as 
\begin{equation}\label{EnergyDensity}
        \mathcal{E}= \frac{1}{2}\ \widehat{\omega}_0^2\ a_0^2\ \widehat{\eta}_0^2.
\end{equation}
If we now integrate \eqref{EnergyDensity} over the real axis, the total energy stored in this vibration mode will be 
\begin{equation}\label{TotalEnergy}
        E=\int^{\infty}_{-\infty} \mathcal{E}\ dx=\frac{1}{2}\ \widehat{\omega}_0^2\ a_0^2\ \widehat{C}_{0},
\end{equation}
where $\widehat{C}_{0}$ is defined in \eqref{CDj}.

On the other hand, the total average power emitted in a period by the plane wave $\eta=A \sin(\omega t- q x+\delta)$ in both parts of the real axis is $\langle P \rangle =\frac{d E}{d t}=- A^2\, \omega\, q$.
For this mode there is only one radiation term for $\kappa>6$, its frequency being $2\widehat{\omega}_0$.
If we rewrite the amplitude described by \eqref{FinalRadiationFirstField} as
\begin{equation}
    \overline{A}_{2\widehat{\omega}_0}=a_0^2\  \overline{A}'_{2\widehat{\omega}_0}
\end{equation}
then, the radiated power emitted by the wobbling kink is
\begin{equation}\label{RadiatedEnergy}
    \langle P\rangle=\frac{d E}{d t}=- (a_0^2\,  \overline{A}'_{2\widehat{\omega}_0})^2\, (2\, \widehat{\omega}_0)\, \overline{q},
\end{equation}
 where $\overline{q}=\sqrt{4\, \widehat{\omega}_0^2-4}$. 
Taking into account the equations \eqref{TotalEnergy} and \eqref{RadiatedEnergy}, we arrive at the differential equation
\begin{equation}
        \frac{1}{2}\, \widehat{\omega}_0^2\  \widehat{C}_{0}\ \frac{d a_0^2(t)}{ d t}\approx- 2\widehat{\omega}_0\, \overline{A}_{2\widehat{\omega}_0}'^{2}\,  \overline{q}\, a_0^4(t), 
\end{equation}
whose solution is 
\begin{equation}\label{DecayLawMode0}
             a_0(t)\approx\frac{a_0(0)}{\sqrt{1+ t\left(\dfrac{4\, \overline{q}\, a_0(0)^2\, 
             \overline{A}_{2\widehat{\omega}_0}'^{2} }{\widehat{C}_{0}\, \widehat{\omega}_0}\right)}} .
\end{equation}
For $\kappa>6$, $a_0(0)=0.2$ and $800<t<1000$, \eqref{DecayLawMode0} predicts that $a_0\approx 0.05-0.04$, which is the same range of values obtained through numerical simulations in Section~\ref{Section4.1} and is represented in the second drawing of Figure~\ref{Fig:Mode0rthogonalAmplitude}.
A completely similar calculation can be done for higher modes, but in these cases the radiation amplitudes are much smaller than for $\widehat{\eta}_{D,0}$, which implies that the decay in $a_0(t)$ will be almost insignificant.

\section{Concluding remarks}

Kink solutions in the $\phi^4$ model have been widely used in the literature to explain numerous natural phenomena whose origin is based on the presence of nonlinear terms in the model that describes the physical system. 
This model implies only a scalar field that can describe a given physical quantity. In this work we propose the study of a more general system that involves two scalar fields. The proposed system is a natural generalization of the $\phi^4$ model with two copies of its potential coupled with a cross term of the type $\kappa \phi^2 \psi^2$. 
This model, likethe $\phi^4$ model, has kink solutions, which can now be perturbed by both longitudinal and orthogonal fluctuations, giving rise to new types of wobblers. It is important to study the evolution of these new solutions (which can appear spontaneously  in any physical system through phase transitions or a thermal bath) since they can critically affect the dynamics of the system and give rise to new behaviors.

Throughout this paper, we have studied in detail the behavior of a wobbling kink and how the excited eigenmode couples with the rest of the shape modes in the context of the aforementioned  two-component scalar field theory.
For this case, it was found that the shape mode structure depends on the value of the coupling constant between both field components. 
In addition to this, we also found that the wobbler emits radiation with frequency $2\widehat{\omega}_j$ in the longitudinal channel, and with frequencies $3\widehat{\omega}_j$ and $\widehat{\omega}_j+\overline{\omega}$ in the orthogonal channel, where $\widehat{\omega}_j$ and $\overline{\omega}$ are the frequencies associated with the  orthogonal shape mode and  longitudinal mode triggered. 
This differs from what is known for the $\phi^4$ model, where the wobbling kinks only have one radiation term that emits with frequency $2\overline{\omega}$. 
The value of these amplitudes also depends on the coupling constant $\kappa$, a parameter that also determines whether a frequency is embedded in the continuous spectrum and, therefore, whether the kink is capable of emitting radiation with that frequency. 

We can see a clear example of this when $\widehat{\eta}_{D,0}$ is activated since in this case, if $\kappa<6$, there is no frequency radiation $2\widehat{\omega}_0$, in contrast to what happens when higher orthogonal modes are excited. Another example can be found when analyzing the frequencies $\overline{\omega}+\widehat{\omega}_j$, since these terms are only included in the continuous part of the spectrum of the orthogonal channel for a range of values of $ \kappa$.

In addition to what has already been mentioned, the coupling mechanism between shape modes has also been studied, which allowed us to find that the triggered shape mode also couples with shape modes that have the same parity and not only with the longitudinal one.
We could also observe some divergences in both  shape mode amplitudes and radiation amplitudes, due to resonances between frequencies in the vibrational spectrum of the small second-order kink fluctuation.

Another notable phenomenon that appears among the results of this work is the decay in the wobbling amplitude due to the loss of energy in the form of radiation.
This energy loss was of great importance when the first orthogonal mode was studied, since when this mode is triggered the emitted radiation is much greater than when higher eigenmodes are considered.

As a future line of research that serves as a natural continuation of this work, we consider the possibility of using the techniques presented here to study the behavior of vortices in the abelian Higgs model that have been triggered by one of its excited states, something that may be physically very relevant, given the ubiquity of this type of systems in various physical applications. Work in this direction is in progress.

\section*{Declaration of Competing Interest} 

The authors declare no competing interests.


\section*{Acknowledgments}

DMC acknowledges financial support from the European Social Fund, the Operational Programme of Junta de Castilla y Leon and the regional Ministry of Education.
This research was supported by Spanish MCIN with funding from European Union Next Generation EU (PRTRC17.I1) and Consejeria de Educacion from JCyL through QCAYLE project, as well as MCIN projects PID2020-113406GB-I00 and RED2022-134301-T.

\bigskip



\begin{thebibliography}{99}

\bibitem{Bishop1978} 
A.R. Bishop and T. Schneider, \textit{Solitons and Condensed Matter Physics} (Springer-Verlag,  1978).


\bibitem{Rajaraman1982} 
R. Rajaraman, \textit{Solitons and Instantons} (North-Holland Personal Library, 1982).


\bibitem{Dauxois2006} 
T. Dauxois and M. Peyrard, \textit{Physics of Solitons} (Cambridge University Press, 2006).


\bibitem{CuevasMaraver2014}  
J. Cuevas-Maraver, P.G. Kevrekidis and  F.William, \textit{The sine-Gordon Model and its Applications} (Springer, 2014).


\bibitem{Kibble1976}  
T.W.B. Kibble, \textit{Topology of Cosmic Domains and Strings},  J. Phys. A: Math. Gen. \textbf{9},  1387 (1976).


\bibitem{Vachaspati2006} 
T. Vachaspati, \textit{Kinks and Domain walls: An Introduction to classical and quantum solitons} (Cambridge University Press,  2006).


\bibitem{Vilenkin1994} 
A. Vilenkin and E.P.S. Shellard,   \textit{Cosmic Strings and other Topological Defects} (Cambridge University Press, 2000).


\bibitem{Buzea1998} 
C. Buzea and T. Yamashita, \textit{Generalization of the Kink Solution for Superconductors with Large Penetration Depths in the Ginzburg-Landau Formalism}, Chaos Solitons Fractals \textbf{10}, 1529 (1999).


\bibitem{Dymarsky2021} 
A. Dymarsky and A. Shapere, \textit{Solutions of modular bootstrap constraints from quantum codes}, Phys. Rev. Lett. \textbf{126}, 161602 (2021) .

\bibitem{Buican2023} 
M. Buican, A. Dymarsky and R. Radhakrishnan, \textit{Quantum codes, CFTs, and defects}, J. High Energ. Phys. \textbf{2023}, 17 (2023).    


\bibitem{Goodman2005}  
R.H.  Goodman and R.  Haberman, \textit{Kink-Antikink Collisions in the $\phi^4$ Equation: The n-Bounce Resonance and the Separatrix Map}, SIAM J. Appl. Dyn. Syst. \textbf{4}, 1195 (2005).


\bibitem{AlonsoIzquierdo2021b} 
A. Alonso-Izquierdo, L.M. Nieto and  J. Queiroga-Nunes, \textit{Asymmetric scattering between kinks and wobblers}, Commun. Nonlinear Sci. Numer. Simul. \textbf{107}, 106183 (2021).


\bibitem{AlonsoIzquierdo2021c}
A. Alonso-Izquierdo, L.M. Nieto and  J. Queiroga-Nunes, \textit{Scattering between wobbling kinks}, Phys. Rev. D \textbf{103}, 045003 (2021). 


\bibitem{Mohammadi2022} 
M. Mohammadi and E. Momeni, \textit{Scattering of kinks in the $B\phi^4$ model}, Chaos Solitons Fractals \textbf{165}, 112834 (2022).



\bibitem{Kevrekidis2019} 
P.G. Kevrekidis and R.H. Goodman, \textit{Four Decades of Kink Interactions in Nonlinear Klein-Gordon Models: A Crucial Typo, Recent Developments and the Challenges Ahead}, arXiv:1909.03128v1 (2019).






\bibitem{Springer2019} 
P.G. Kevrekidis and J. Cuevas-Maraver, \textit{A Dynamical Perspective on the $\phi^4$ Model}. Nonlinear Systems and Complexity, vol 26 (Springer,  2019).


\bibitem{Dorey2011} 
P. Dorey, K. Mersh, T. Romanczukiewicz, and Y. Shnir, \textit{Kink-Antikink Collisions in the $\phi^6$ Model}, Phys. Rev. Lett. \textbf{107}, 091602 (2011).


\bibitem{Marjaneh2017}
A.M. Marjaneh, V.A. Gani, D. Saadatmand, S.V. Dmitriev and K. Javidan,  \textit{Multi-kink collisions in the $\phi^6$ model}, J. High Energy Phys. \textbf{2017}, 28 (2017).


\bibitem{Gani2014} 
V.A. Gani, A.E. Kudryavtsev  and M.A. Lizunova, \textit{Kink interactions in the (1+1)-dimensional $\phi^6$ model}, Phys. Rev. D \textbf{89}, 125009 (2014).




\bibitem{Belendryasova2018} 
E. Belendryasova  and  V.A. Gani, \textit{Scattering of the $\phi^8$ kinks with power-law asymptotics}, Commun. Nonlinear Sci. Numer. Simul. \textbf{67}, 414 (2018).


\bibitem{Bazeia2023} 
D.  Bazeia, J.G.F. Campos and  A. Mohammadi, \textit{Kink-antikink collisions in the $\phi^8$ model: short-range to long-range journey},  	arXiv:2303.12482  (2023)


\bibitem{Gani2015} 
V.A. Gani, V. Lensky and M.A. Lizunova, \textit{Kink excitation spectra in the (1+1)-dimensional $\phi^8$ model}, J. High Energy Phys. \textbf{2015}, 147 (2015).


\bibitem{Katsura2013}
H. Katsura, \textit{Composite-kink solutions of coupled nonlinear wave equations}, Phys. Rev. D \textbf{89}, 085019  (2013).


\bibitem{Bazeia1995}  
D. Bazeia, M.J. dos Santos and R.F. Ribeiro, \textit{Solitons in systems of coupled scalar fields},     Phys. Lett. A \textbf{208},  84 (1995).


\bibitem{Shifman1998}
M.A. Shifman and M.B. Voloshin,  \textit{Degenerate domain wall solutions in supersymmetric theories},  Phys. Rev. D \textbf{57}, 2590 (1998).


\bibitem{AlonsoIzquierdo2019} 
A. Alonso-Izquierdo, \textit{Kink dynamics in the MSTB model}, Phys. Scr. \textbf{94}, 085302(2019)


\bibitem{Ashcroft2016}   
J. Ashcroft, M. Eto, M.  Haberichter, M. Nitta and M.B. Paranjape, \textit{Head butting sheep: kink collisions in the presence of false vacua}, J. Phys. A \textbf{49} , 365203 (2016).


\bibitem{Aguirre2020} 
A.R. Aguirre and E. S. Souza,  \textit{Extended multi-scalar field theories in (1 + 1) dimensions},    Eur. Phys. J. C \textbf{80}, 1143 (2020).


\bibitem{AlonsoIzquierdo2002}    
A. Alonso-Izquierdo, M.A. Gonzalez-Leon and J. Mateos-Guilarte, \textit{Kink variety in systems of two coupled scalar fields in two space-time dimensions},     Phys. Rev. D \textbf{65}, 085012 (2002).


\bibitem{Halavanau2012} 
A. Halavanau, T. Romanczukiewicz and Y.M. Shnir, \textit{Resonance structures in coupled two-component $\phi^4$ model}, Phys. Rev. D \textbf{86}, 085027 (2012).


\bibitem{AlonsoIzquierdo2021}
A. Alonso-Izquierdo, M.A. Gonz\'alez Le\'on, J. Mart\'in Vaquero and M. de la Torre Mayado, \textit{Kink scattering in a generalized Wess-Zumino model}, Commun. Nonlinear. Sci. Numer. Simulat. \textbf{103},  106011 (2021).


\bibitem{AlonsoIzquierdo2023}
A. Alonso-Izquierdo, D. Migu\'elez-Caballero, L.M. Nieto and J. Queiroga-Nunes, \textit{Wobbling kinks in a two-component scalar field theory:
Interaction between shape modes}, Physica D \textbf{443}, 133590 (2023).



\bibitem{Blinov2022}
P.A. Blinov, T.V. Gani, A.A. Malnev, V.A. Gani  and V.B. Sherstyukov,  \textit{Kinks in higher-order polynomial models}, 
Chaos Solitons Fractals \textbf{165}, 112805 (2022).


\bibitem{Christov2021} 
I.C. Christov, R.J. Decker, A. Demirkaya, V.A. Gani, P.G. Kevrekidis and A.  Saxen, \textit{Kink-antikink collisions and multi-bounce resonance windows in higher-order field theories}, Commun. Nonlinear Sci. Numer. Simul. \textbf{97}, 105748   (2021).



\bibitem{Askari2020} 
A. Askari, et al,  \textit{Collision of $\phi^4$ kinks free of the Peierls-Nabarro barrier in the regime of strong discreteness}, Chaos Solitons Fractals \textbf{138}, 109854 (2020),


\bibitem{Takyi2023} 
I. Takyi, S. Gyampoh, B. Barnes, J. Ackora-Prah and G.A. Okyere, \textit{Kink collision in the noncanonical $\phi^6$ model: A model with localized inner structures},  Results Phys.  \textbf{44}, 106197 (2023).


\bibitem{Mohammadi2021} 
M. Mohammadi and R. Dehghani, \textit{Kink-antikink collisions in the periodic $\phi^4$ model}, Commun. Nonlinear Sci. Numer. Simul. \textbf{94}, 105575 (2021).



\bibitem{Sugiyama1978}
T. Sugiyama, \textit{Kink-Antikink Collisions in the Two-Dimensional $\phi^4$ Model},     Prog. Theor. Phys. \textbf{61},  1550 (1978).


\bibitem{NavarroObregon2023}
S. Navarro-Obreg\'on,  L.M. Nieto  and J.M. Queiruga \textit{Inclusion of radiation in the CCM approach of the $\phi^4$ model},  	arXiv:2305.00497 (2023).


\bibitem{Manton2004} 
N.S. Manton and P. Sutcliffe, \textit{Topological Solitons}  (Cambridge University Press, 2004).


\bibitem{Adam2022}   
C. Adam, N.S. Manton, K. Oles, T. Romanczukiewicz and  A. Wereszczynski, \textit{Relativistic Moduli Space for Kink Collisions},      Phys. Rev. D \textbf{105}, 065012 (2022).


\bibitem{Adam2023}    
C. Adam, D. Ciurla, K. Oles, T. Romanczukiewicz and  A. Wereszczynski \textit{Relativistic Moduli Space and critical velocity in kink collisions}, 	arXiv:2304.14076  (2023). 


\bibitem{Takyi2016}   
I. Takyi and  H. Weigel, \textit{ Collective Coordinates in One–Dimensional Soliton Models Revisited}, Phys. Rev. D \textbf{94}, 085008 (2016).










\bibitem{Flugge1971} 
S. Fl\"ugge, \textit{Practical Quantum Mechanics} (Springer, 1971).


\bibitem{Morse1953} 
P. Morse and  H. Feshbach, \textit{Methods of Theoretical Physics} (McGraw-Hill Book Company, 1953).


\bibitem{Morse1933}
P. Morse and N. Rosen, \textit{On the Vibrations of Polyatomic Molecules}, Phys. Rev. \textbf{42}, 210 (1933).


\bibitem{Shnir2018} 
Y.M. Shnir, \textit{Topological and non-topological solitons in scalar field theories} (Cambridge University Press, 2018).


\bibitem{Manton1997}
N.S. Manton and  H. Merabet, \textit{Kinks-gradient flow and dynamics}, Nonlinearity \textbf{10},  3 (1997).


\bibitem{Barashenkov2009} 
I.V. Barashenkov and O.F. Oxtoby, \textit{Wobbling kinks in $\phi^4$ theory}, Phys. Rev. E \textbf{80},  026608 (2009).


\bibitem{Barashenkov2019} 
I.V. Barashenkov, \textit{The Continuing Story of the Wobbling Kink}. In: P.G.  Kevrekidis and J. Cuevas-Maraver (eds) \textit{A Dynamical Perspective on the $\phi^4$ Model}. Nonlinear Systems and Complexity, vol 26 (Springer, 2019), pp. 187--212.



\bibitem{NIST2010}
F.W.J. Olver, D. W. Lozier, R. F. Boisvert and C. W. Clark, \textit{NIST Handbook of mathematical functions} (Cambridge University Press, 2010).





\end{thebibliography}
\end{document}